\pdfoutput=1
\documentclass[12pt]{article}
\usepackage{epsf,amsmath,amssymb,graphicx,dcolumn}
\usepackage{scalefnt,cite}
\usepackage[hyperfootnotes=false]{hyperref}
\usepackage{color,soul}
\usepackage{listings,booktabs}
\usepackage[utf8]{inputenc}
\usepackage[hang,flushmargin]{footmisc}

\newcommand{\citere}[1]{Ref.\,\cite{#1}}
\newcommand{\citeres}[1]{Refs.\,\cite{#1}}

\newcommand{\abbrev}{\scalefont{.9}}
\newcommand{\eqn}[1]{Eq.\,(\ref{#1})}
\newcommand{\fig}[1]{Fig.\,\ref{#1}}
\newcommand{\tab}[1]{Tab.\,\ref{#1}}
\newcommand{\sct}[1]{Section~\ref{#1}}
\newcommand{\appref}[1]{Appendix~\ref{#1}}

\newcommand{\epa}{{\abbrev EPA}}
\newcommand{\eft}{{\abbrev EFT}}

\newcommand{\sm}{{\abbrev SM}}
\newcommand{\thdm}{{\abbrev 2HDM}}
\newcommand{\mssm}{{\abbrev MSSM}}
\newcommand{\hmssm}{{\abbrev hMSSM}}
\newcommand{\susy}{{\abbrev SUSY}}

\newcommand{\cp}{{\abbrev $\mathcal{CP}$}}
\newcommand{\lo}{{\abbrev LO}}
\newcommand{\nlo}{{\abbrev NLO}}

\newcommand{\GeV}{{\rm GeV}}

\newcommand{\s}{\newline \vspace*{-3.5mm}}
\newcommand{\beq}{\begin{equation}}
\newcommand{\eeq}{\end{equation}}
\newcommand{\bea}{\begin{eqnarray}}
\newcommand{\eea}{\end{eqnarray}}
\newcommand{\msto}{m_{\tilde t_1}}
\newcommand{\mstt}{m_{\tilde t_2}}
\newcommand{\msta}{m_{\tilde t_a}}
\newcommand{\mstb}{m_{\tilde t_b}}
\newcommand{\mstc}{m_{\tilde t_c}}

\begin{document}

\begin{flushleft}
\end{flushleft}
\vspace*{-1.35cm}
\begin{flushright}
{\tt KA-TP-29-2018},
{\tt PSI-PR-18-12 }
\end{flushright}

\long\def\symbolfootnote[#1]#2{\begingroup%
\def\thefootnote{\fnsymbol{footnote}}\footnote[#1]{#2}\endgroup}

\vspace{0.1cm}

\begin{center}
\Large\bf\boldmath
The \hmssm{} approach for Higgs self-couplings revisited
\unboldmath
\end{center}
\vspace{0.05cm}
\begin{center}
Stefan Liebler$^a$,
Margarete Mühlleitner$^a$,
Michael Spira$^b$,
Maximilian Stadelmaier$^{a,c}$\symbolfootnote[0]{\noindent Electronic addresses:
stefan.liebler@kit.edu, milada.muehlleitner@kit.edu,\\
michael.spira@psi.ch, maximilian.stadelmaier@kit.edu.}\\[0.4cm]
{\small
{\sl${}^a$Institute for Theoretical Physics, Karlsruhe Institute of Technology,
D-76131 Karlsruhe, Germany}\\[0.2em]
{\sl${}^b$Paul Scherrer Institut, CH-5232 Villigen PSI, Switzerland}\\[0.2em]
{\sl${}^c$Institute for Nuclear Physics, Karlsruhe Institute of Technology,
D-76344 Karlsruhe, Germany}
}
\end{center}

\begin{abstract}
\noindent
We compare the decay of the heavy Higgs boson into two \sm{}-like Higgs
bosons, $H\to hh$, calculated in a Feynman-diagrammatic approach at the one-loop level
based on the one hand on the full effective potential involving the top quark and stops
in the Minimal Supersymmetric Standard Model (\mssm{})
accompanied by the matched Two-Higgs-Doublet Model (\thdm{}) as its low-energy limit
and on the other hand on the \hmssm{} approximation.  We identify
missing contributions due to the top quark in the Higgs self-couplings of
the \hmssm{}, that -- when taken into account -- lead to a good
agreement between the \hmssm{} and a full \mssm{} calculation, at least
in the limit of the Higgsino mass parameter~$\mu$
being small compared to the stop spectrum. We also
thoroughly analyze momentum-dependent and kinetic
corrections intrinsic to the
Feynman-diagrammatic approach and the matching to the effective Lagrangian, respectively,
for both our calculation in the \mssm{} and the \hmssm{} and for the latter suggest to include 
additional corrections from the top quark, which are
independent of the unknown supersymmetric spectrum.
\end{abstract}

\section{Introduction}

The discovery of the Higgs boson with a mass of ($125.09\pm0.24$)~GeV
\cite{Aad:2015zhl} in 2012 by the LHC experiments ATLAS
\cite{Aad:2012tfa} and CMS \cite{Chatrchyan:2012ufa} has marked a
milestone for particle physics. While this structurally completes the
Standard Model (\sm{}) the \sm{} itself leaves open many questions that
require extensions of the model. The \sm{} is therefore considered as an
effective low-energy description of a more complete model valid at
high-energy scales.  Since the discovered Higgs boson behaves very
\sm{}-like any such beyond-the-\sm{} theory has to contain a \sm{}-like
Higgs boson with a mass of about $125$\,GeV.  \s

The Higgs sector of the Minimal Supersymmetric extension of the \sm{}
(\mssm{}) \cite{Golfand:1971iw, Volkov:1973ix, Wess:1974tw, Fayet:1974pd,
Fayet:1977yc, Fayet:1976cr, Nilles:1983ge, Haber:1984rc, Sohnius:1985qm,
Gunion:1984yn, Gunion:1986nh, Gunion:1989we, Djouadi:2005gj} consists of
two complex Higgs doublets to ensure supersymmetry and the cancellation
of anomalies. After electroweak symmetry breaking its Higgs
sector contains five physical Higgs bosons, two neutral \cp{}-even bosons,
$h,H$, one neutral \cp{}-odd boson, $A$, and a charged Higgs pair, $H^\pm$.
The tree-level Higgs sector can be described by two parameters, usually
chosen to be the mass of the \cp{}-odd Higgs boson, $M_A$, and the ratio of the
two vacuum expectation values of the two Higgs doublets,
$\tan\beta=v_2/v_1$, in the case of real supersymmetric parameters.
Supersymmetry restricts the tree-level mass of the lightest \cp{}-even
scalar $h$ to values below the $Z$ boson mass~$M_Z$.  This constraint is
relaxed by the inclusion of radiative corrections in the Higgs sector
that can shift its value to the measured $125$\,GeV.  The dominant
corrections originate from third generation quark/squark loops.
Depending on the parameter choices, the squark masses must be quite
large in order to match the observed Higgs mass value for small values
of $\tan\beta$. Moreover, in a significant part of the \mssm{} parameter
space the limits on the squark masses are pushed into the TeV range by
the unsuccessful LHC searches for supersymmetric (\susy{}) particles so
far. The loop-corrected Higgs sector depends on many \susy{} parameters so
that the investigation of the \mssm{} parameter space becomes a
complicated task.  This triggered the introduction of benchmark
scenarios that are used by the experimental collaborations for the
interpretation of their results.  Among these, the \hmssm{} presented in
\citeres{Djouadi:2013uqa, Djouadi:2013vqa, Maiani:2013hud, Djouadi:2015jea}
exploits the fact that the dominant corrections to the lightest \cp{}-even
Higgs mass and the mixing parameters that enter the Higgs
couplings have a common origin and that the dominant corrections stem
from the top-quark and its supersymmetric partners, the stops.  \s

In the \hmssm{} the measured Higgs mass value $M_h$ is taken as an input
parameter in addition to $M_A$ and $\tan\beta$. This removes the explicit
dependence of the Higgs sector on other \susy{} parameters through the
radiative corrections. In its region of applicability, the \hmssm{}
approach has been shown to describe the \mssm{} Higgs mass spectrum and
mixing angle $\alpha$ of the \cp{}-even sector very well
\cite{Bagnaschi:2015hka, Djouadi:2015jea, Lee:2015uza}. In particular,
it allows to probe the low $\tan\beta$ regime where a very high \susy{}
scale is required for the radiative corrections to be large enough to
achieve $125$\,GeV for the light \cp{}-even Higgs mass. The Higgs
self-couplings that are related to the Higgs masses through the
Higgs potential are also affected by large radiative corrections. In
order to make reliable predictions, the large logarithms that appear in
the corrections in case of very large \susy{} masses have to be resummed
using effective field theory (\eft{}) methods. In physical processes
containing the trilinear Higgs self-couplings, like Higgs decays into a
pair of lighter Higgs bosons, momentum-dependent corrections to the
vertex and to the kinetic factors can become important. These are not
taken into account in the \eft{} approach, however, and have to be computed
through a diagrammatic fixed-order calculation. \s

In this paper, we revisit the \hmssm{} approach with focus on the
Higgs-to-Higgs decay of the heavier $H$ into two \sm{}-like Higgs bosons, 
$H\to hh$. We compute the decay at next-to-leading order (\nlo{}) taking
into account the dominant radiative corrections from the top-quark and stop
sector. The calculation is performed in an effective low-energy
\thdm{} with \mssm{}-like quartic couplings
that are properly matched to the \mssm{}
and in the \mssm{} itself. In both cases the calculation is performed in the
Feynman-diagrammatic approach thus including
momentum-dependent corrections. Moreover, radiative corrections to the
Higgs self-couplings from the top-quark contributions in the \thdm{}, and the top-quark
and stop contributions in the \mssm{}, are taken into account through
effective couplings. By choosing appropriate counterterms according to
the low-energy limit, double counting is avoided when including the
diagrammatic NLO corrections. By plugging in the effective trilinear
Higgs self-coupling of the \hmssm{} and comparing with the full \mssm{}
result, we are able to disentangle the deviations due to the \hmssm{}
approximation of the coupling from those originating from
momentum-dependent contributions. In this way, we are able to properly
dissect the Higgs self-coupling of the effective \hmssm{} approximation and
to propose improvements that allow to better approximate the full
result. It turns out that the bulk of the improvement does not introduce
additional parameters so that it is appropriate to dub it ``improved
\hmssm{}''. \s

The outline of the paper is as follows. In
Section~\ref{sec:mssmhiggssector} we introduce the \mssm{}, the \thdm{}
and the \hmssm{} approach and define our notation. In particular in
Subsection~\ref{sec:effpot} we discuss the effective potential as part
of the low-energy effective Lagrangian, while
Subsection~\ref{sec:improve} presents our proposed improvement of the
\hmssm{} approach.  Section~\ref{sec:calc} contains the explicit
computation of the \nlo{} decay width for the decay $H \to hh$
with effective couplings.
Section~\ref{sec:numerical} is dedicated to the presentation of our
numerical results.  Our conclusions are given in
Section~\ref{sec:concl}.

\section{The MSSM Higgs sector as an effective 2HDM \label{sec:mssmhiggssector}}

In this section we first introduce our notation of the \mssm{} Higgs
sector.  If the supersymmetric spectrum is heavy, the \mssm{} Higgs
sector can be understood as an effective (properly matched) low-energy
\thdm{}, where all heavy
(s)particles are integrated out. This approach allows to resum
logarithms as they appear e.g. in the derivation of the Higgs
masses at loop-level.  In this section we use the Effective
Potential Approach (\epa{})~\cite{Coleman:1973jx,Jackiw:1974cv}
for the matching,
define the relevant effective potential, kinetic
corrections in the effective Lagrangian and provide results for loop
corrections to the Higgs masses and the Higgs self-couplings at
order $\mathcal{O}(\alpha_t)$, where $\alpha_t=y_t^2/(4\pi)$ with the
top-Yukawa coupling $y_t = \sqrt{2} m_t/v$ ($v\approx 246.22$\,GeV is the \sm{}
vacuum expectation value and $m_t$ the top-quark mass.).  We explain the
\hmssm{} approach and identify terms that are missing in the Higgs
self-couplings of the \hmssm{} approach.

\subsection{The MSSM Higgs and squark sectors at tree-level}
Supersymmetry requires the introduction of at least two complex Higgs
doublets, which is realized in the minimal \susy{} version, the \mssm{}. In
addition, the Adler-Bell-Jackiw anomaly
contributions~\cite{Adler:1969gk,Bell:1969ts}
due to the Higgsino doublet states cancel each other thanks to their
opposite hypercharges. The two doublets $H_u$ and $H_d$ with hypercharge
$Y=1$ and $-1$, respectively, can be expressed in terms of the charged
and neutral components $\phi_i^\pm$ and $\phi_i^0$ ($i=u,d$), as  
\beq
H_d = \left( \begin{array}{c} \phi_d^{0*} \\- \phi_d^- \end{array}
\right) \quad \mbox{and} \quad
H_u = \left( \begin{array}{c} \phi_u^{+} \\ \phi_u^0 \end{array}
\right) \,.
\eeq
The tree-level potential for the \mssm{} Higgs fields is derived from the
$F$- and $D$-term contributions and the soft-\susy{} breaking
Lagrangian. The most general supersymmetric potential for two Higgs
doublets at tree level reads
\bea
\label{eq:V0mssm}
V^{\text{\lo{}}}_{\text{\mssm{}}} &=& (m_{H_d}^2 + \mu^2) |H_d|^2 + (m_{H_u}^2 + \mu^2) |H_u|^2 
- B\mu \epsilon_{ij} (H_d^i H_u^j + h.c.) \nonumber \\
&& + \frac{g^2+{g'}^{2}}{8} (|H_d|^2- |H_u|^2)^2
+ \frac{g^2}{2} |H_d^* H_u|^2 \;,
\eea
where $\epsilon_{12}= - \epsilon_{21} = 1$ and $m_{H_d}^2$, $m_{H_u}^2$
and $B \mu$ denote the corresponding soft-\susy{} breaking mass parameters.
The SU$(2)_L$
and U$(1)_Y$ gauge couplings are given by $g$ and $g'$, respectively.
After electroweak symmetry breaking we expand the neutral
fields around the vacuum expectation values (VEVs) according to 
\bea
\label{eq:neutralfieldsexpand}
\phi_d^0  =  \frac{1}{\sqrt{2}} \left( v_d + \sigma_d + i\xi_d \right)\,,\qquad
\phi_u^0  =  \frac{1}{\sqrt{2}} \left( v_u + \sigma_u + i\xi_u \right)\,.
\eea
The ratio of the VEVs $v_u$ and $v_d$ is defined as $\tan\beta = \frac{v_u}{v_d}$ 
while obeying the sum rule $v_u^2 + v_d^2 = v^2$.
Throughout our work we fix $m_{H_d}^2$ and $m_{H_u}^2$ through the
radiatively corrected tadpole equations.
Rotating from the gauge eigenstates $\xi_d$ and $\xi_u$ to the mass eigenstates
by the mixing angle $\beta$ in the \cp{}-odd sector
yields a massless Goldstone boson~$G^0$ and the \cp{}-odd Higgs boson~$A$ with
mass
\beq
\label{eq:MAmssm}
M_A^2 = \frac{2 B\mu}{\sin 2\beta } \;.
\eeq
We define the $Z$ boson mass $M_Z=\tfrac{1}{2}\sqrt{g^2+g'^2}~v$, such that
the tree-level mass matrix of the \cp{}-even sector in
the gauge eigenstates $\sigma_d$ and $\sigma_u$ takes the form
\bea
\label{eq:treemassmssm}
\mathcal{M}_{\text{tree}}^2=\begin{pmatrix}\mathcal{M}_{dd}^2&\mathcal{M}_{du}^2\\\mathcal{M}_{du}^2&\mathcal{M}_{uu}^2\end{pmatrix}
=\begin{pmatrix}M_A^2s_\beta^2+M_Z^2c_\beta^2 & -(M_A^2+M_Z^2)s_\beta c_\beta\\
-(M_A^2+M_Z^2)s_\beta c_\beta & M_A^2c_\beta^2+M_Z^2s_\beta^2\end{pmatrix}\,.
\eea
Where appropriate we use $s_x, c_x$ and $t_x$ as abbreviations
for $\sin(x), \cos(x)$ and $\tan(x)$, respectively.
The matrix is diagonalized through a rotation by the \cp{}-even mixing angle $\alpha$,
which is given by
\bea
\tan 2\alpha=-\frac{2\mathcal{M}_{du}^2}{\mathcal{M}_{uu}^2-\mathcal{M}_{dd}^2}\,.
\label{eq:treealpha}
\eea
This rotation results in two mass eigenstates, the \cp{}-even Higgs bosons~$h$ and $H$, with an upper bound of 
\beq
M_h^2 \le M_Z^2 \cos^2 2\beta
\eeq
on the tree-level mass of the lightest \cp{}-even Higgs boson. The
dominant radiative corrections to the Higgs mass originate from
the top-quark and stop loops, which we subsequently discuss in the effective potential approach.
The Higgs self-couplings in the mass eigenstates are given by the relations
\begin{align}
\label{eq:LOlambdas}
\lambda_{hhh}  &=  3 \frac{M_Z^2}{v}c_{2\alpha} s_{\alpha+\beta}\,,\qquad
&\lambda_{Hhh}  &=  \frac{M_Z^2}{v} ( 2 s_{2\alpha}s_{\alpha+\beta} - c_{2\alpha} c_{\alpha+\beta})\,,\nonumber \\
\lambda_{HHh}  &=  \frac{M_Z^2}{v} (- 2 s_{2\alpha} c_{\alpha+\beta} - c_{2\alpha} s_{\alpha+\beta})\,,\qquad
&\lambda_{HHH}  &=  3\frac{M_Z^2}{v} c_{2\alpha} c_{\alpha+\beta}\,, \nonumber \\
\lambda_{hAA}  &=  \frac{M_Z^2}{v} c_{2\beta} s_{\alpha+\beta}\,,\qquad
&\lambda_{HAA}  &=  - \frac{M_Z^2}{v} c_{2\beta} c_{\alpha+\beta}\,.
\end{align}
The mass matrix of the stop sector using left- and right-handed stops
$\tilde t_L$ and $\tilde t_R$ is given by
\bea
\label{eq:squarkmassmatrix}
\mathcal{M}^2_{\tilde t}=\begin{pmatrix}M_{\tilde t_L}^2+m_t^2& m_tX_t\\m_tX_t & M_{\tilde t_R}^2+m_t^2\end{pmatrix}\,,
\eea
where $M_{\tilde t_L}$ and $M_{\tilde t_R}$ are the left- and right-handed soft-\susy{} breaking
mass terms, respectively.
$M_{\tilde t_L}$ equals the soft-\susy{} breaking mass term $M_{\tilde Q_L}$
of the doublet of the third generation squarks. 
The stop-mixing parameter $X_t$ is defined through $X_t=A_t-\mu/t_\beta$
involving the trilinear soft-\susy{} breaking stop parameter $A_t$ and the $\mu$-term, which
has already been part of the tree-level potential in \eqn{eq:V0mssm}.
In the subsequent derivation of the effective potential we work in the gaugeless limit,
which is why we also omit $D$-terms proportional to $M_Z^2$ in the stop mass matrix. Its diagonalization
yields the stop masses $\msto$ and $\mstt$, which for $M_S:=M_{\tilde t_L}=M_{\tilde t_R}$
are given by $\msto^2=M_S^2+m_t^2-m_t|X_t|$ and $\mstt^2=M_S^2+m_t^2+m_t|X_t|$.

\subsection{The 2HDM Higgs sector at tree-level}
Being the low-energy limit of the \mssm{}, we work only in the type-II
\thdm{}, where $H_1$ and $H_2$ couple to down-type and up-type
quarks, respectively.
The tree-level Higgs potential including only terms that arise
in the \mssm{} at tree-level takes the form
\bea
\label{eq:V02hdm}
V^{\text{\lo{}}}_{\text{\thdm{}}} &=& m_{1}^2  |H_1|^2 + m_{2}^2  |H_2|^2 
- m_3^2 (H_1^\dagger H_2 + h.c.) \nonumber \\
 &&+\frac{\lambda_1}{2}|H_1|^4 + \frac{\lambda_2}{2}|H_2|^4+\lambda_3|H_1|^2|H_2|^2 + \lambda_4 |H_1^\dagger H_2|^2\;.
\eea
We label the fields $H_1$ and $H_2$ instead of $H_d$ and $H_u$ to follow the
standard notation.
Both field conventions follow the \mssm{} field $H_u$,
i.e. $H_i=(\phi_i^+,\phi_i^0)^T$ ($i=1,2$). 
For the relation between the Higgs fields in the \mssm{} and the \thdm{} we also refer to \citere{Beneke:2008wj}.
The parameters $m_1^2$ and $m_2^2$ can again be fixed through the tadpole equations,
such that in a generic \thdm{} the potential has six free parameters, which in the above
$\lambda$-basis are $m_3^2$, $\lambda_1 - \lambda_4$ and the ratio of the VEVs, $\tan\beta$.
The trilinear Higgs self-couplings within the $\lambda$-basis
are given by
\bea\nonumber
\lambda_{hhh}&=&3v(-\lambda_1s_\alpha^3c_\beta+\lambda_2c_\alpha^3s_\beta-\tfrac{1}{2}\lambda_{34}s_{2\alpha}c_{\alpha+\beta})\,,\\\nonumber
\lambda_{Hhh}&=&v[3\lambda_1s_\alpha^2c_\alpha c_\beta +3\lambda_2s_\alpha c_\alpha^2s_\beta+\lambda_{34}(c_{2\alpha}c_{\alpha+\beta}-\tfrac{1}{2}s_{2\alpha}s_{\alpha+\beta})]\,,\\\nonumber
\lambda_{HHh}&=&v[-3\lambda_1s_\alpha c_\alpha^2c_\beta + 3\lambda_2s_\alpha^2 c_\alpha s_\beta+\lambda_{34}(c_{2\alpha}s_{\alpha+\beta}+\tfrac{1}{2}s_{2\alpha}c_{\alpha+\beta})]\,,\\\nonumber
\lambda_{HHH}&=&3v(\lambda_1c_\alpha^3c_\beta+\lambda_2s_\alpha^3s_\beta+\lambda_{34}\tfrac{1}{2}s_{2\alpha}s_{\alpha+\beta})\,,\\\nonumber
\lambda_{hAA}&=&v[-\lambda_1s_\alpha s_\beta^2c_\beta+\lambda_2 c_\alpha s_\beta c_\beta^2+\lambda_{34}(c_\alpha s_\beta^3-s_\alpha c_\beta^3)]\,,\\
\lambda_{HAA}&=&v[\lambda_1c_\alpha s_\beta^2c_\beta+\lambda_2 s_\alpha s_\beta c_\beta^2+\lambda_{34}(c_\alpha c_\beta^3+s_\alpha s_\beta^3) ]\,.
\eea
In the \mssm{} at tree level the couplings $\lambda_i$ are given by
\bea
\lambda_1=\lambda_2=\frac{g^2+g'^2}{4}\,,\quad
\lambda_3=\frac{g^2-g'^2}{4}\,,\quad \lambda_4=-\frac{g^2}{2}\,,\,\text{so that}\quad
\lambda_{34}\equiv\lambda_3+\lambda_4=-\lambda_1=-\lambda_2
\eea
and $m_3^2=B\mu$ which is thus related to $M_A^2$, see \eqn{eq:MAmssm}.

\subsection{The effective 2HDM Higgs sector beyond tree-level}
\label{sec:efflag}
Beyond tree level the masses and couplings of the \thdm{} discussed in the previous section
receive quantum corrections from heavier (s)particles\footnote{We refer
to the top quark as a heavy particle,
since the leading term in a large top-mass expansion yields reliable
approximations for the whole low-energy \thdm{} sector, i.e.~external
momentum-dependent corrections are numerically subleading.} and thus
form an effective low-energy \thdm{}.
This effective model is obtained through a proper matching at the scale,
where the heavier (s)particles are integrated out.
Practically, this matching can be performed by a Feynman-diagrammatic calculation,
in which all squared external momenta $p^2$ are strictly set to zero.
This limit of vanishing momenta defines the \epa{}~\cite{Coleman:1973jx,Jackiw:1974cv}.
This approach includes all one-particle irreducible diagrams and allows to
define an effective \thdm{} as the low-energy limit of the \mssm{}.
Within the rest of this section we will define the effective \thdm{} relevant
for our purposes and later supplement it with
a Feynman-diagrammatic calculation of the decay~$H\to hh$. \s

The \epa{} allows to calculate the corrections to the
potential~$V^{\text{\lo{}}}_{\text{\thdm{}}}(=V^{\text{\lo{}}}_{\text{\mssm{}}})$
in \eqn{eq:V02hdm} and thus defines the effective
potential~$V^{\text{eff}}_{\text{\thdm{}}}$.\footnote{We name it \thdm{} effective potential,
since the \thdm{} is the low-energy theory and we want to separate quark from squark contributions. If both quarks and squarks
are included it is commonly named \mssm{} effective potential.}
A detailed discussion follows in the next section.
However, also the kinetic term of the effective Lagrangian receives
corrections, such that we can decompose the effective
low-energy \thdm{} Lagrangian as
\bea
\label{eq:2hdmlag}
{\cal L}^{\text{eff}}_{\text{\thdm{}}} & = & \sum_{i,j\in \{1,2\}} Z^{\text{eff}}_{ij} (D_\mu H_i)^\dagger D_\mu H_j -
V^{\text{eff}}_{\text{\thdm{}}}\,,
\eea
which involves an additional kinetic matrix $Z^{\text{eff}}$.
The matrix $Z^{\text{eff}}$ can be obtained, as in \citere{Beneke:2008wj},
by an expansion of the (off-)diagonal $H_i H_j~(i,j=1,2)$ two-point
functions in their external momenta.
In our Feynman diagrammatic calculation $Z^{\text{eff}}$ appears as an effective
wave-function renormalization and is named kinetic correction.
We will present its detailed form in \sct{sec:wvandrenorm}. \s

We note that beyond tree level also couplings $\lambda_5$ to $\lambda_7$
are generated, see e.g.~in \citere{Beneke:2008wj}.  Their generation is
connected to a non-vanishing value of $\mu$~\cite{Hall:1993gn}.
However, since we work with physical quantities as the Higgs
self-couplings, masses and mixings derived from the full effective Higgs
potential we do not need their explicit form in this work
since they are taken into account intrinsically.  The \mssm{}
Higgs masses in the \epa{} were first calculated at order
$\mathcal{O}(\alpha_t)$ in
\citeres{Okada:1990vk,Haber:1990aw,Ellis:1990nz,Ellis:1991zd}, followed
by calculations including $\mathcal{O}(\alpha_b)$ and electroweak
corrections in
\citeres{Brignole:1992uf,Chankowski:1992er,Dabelstein:1994hb,Pierce:1996zz}.
In this approach also the large radiative corrections to the Higgs
self-interactions are known at one-loop order ${\cal O}(\alpha_{t,b})$
\cite{Barger:1991ed, Hollik:2001px, Dobado:2002jz} and two-loop order
${\cal O}(\alpha_t\alpha_s)$ \cite{Brucherseifer:2013qva}.  In the
\hmssm{} approach, which is anyhow agnostic to the exact form of the
higher-order corrections, the corrections to the Higgs masses and
the Higgs self-couplings are reexpressed in terms of the light Higgs
mass $M_h$ only, see \sct{sec:hMSSMapproach}.  In both cases, the
proper calculation of the \thdm{} effective potential and the \hmssm{}
approach, we work with an effective \thdm{}.

\subsection{The 2HDM effective potential}
\label{sec:effpot}
As we argued beforehand the application of the \epa{} defines
the Higgs potential of an effective low-energy \thdm{},
which we provide explicitly in this section. We consistently 
distinguish quark from squark contributions, in order to later omit the effect of stops. \s

We are only interested in the $\mathcal{O}(\alpha_t)$ corrections
to the effective potential of the \thdm{} in view of the corrections that
are implicitly taken into account in the \hmssm{} approach.
To stay strictly at $\mathcal{O}(\alpha_t)$ in perturbation theory, we are working in the gaugeless limit and thus drop supersymmetric $D$-term contributions
beyond tree-level. This limit captures the dominant corrections and has the advantage that our discussion of $H\to hh$
remains independent of the renormalization of electroweak parameters,
in particular the vacuum expectation value(s), see \sct{sec:wvandrenorm}.
We also omit corrections from the bottom and sbottom sector.
We split the $\mathcal{O}(\alpha_t)$ corrections to the effective potential in
top- and stop contributions as follows:
\bea
\label{eq:V1mssm}
V^{\text{\nlo{}}}(t) & = & \frac{3}{(4\pi)^2} C_\varepsilon \left\{ \overline m_t^4 \left[
\frac{1}{\varepsilon} + \frac{3}{2} - \log\frac{\overline m_t^2}{Q^2} \right] \right\}\,, \nonumber \\
V^{\text{\nlo{}}}(\tilde t) & = & - \frac{3}{(4\pi)^2} \frac{1}{2} C_\varepsilon \left\{ \overline m_{\tilde t_1}^4 \left[
\frac{1}{\varepsilon} + \frac{3}{2} - \log\frac{\overline m_{\tilde t_1}^2}{Q^2} \right]
+ \overline m_{\tilde t_2}^4 \left[
\frac{1}{\varepsilon} + \frac{3}{2} - \log\frac{\overline m_{\tilde t_2}^2}{Q^2} \right]
\right\}\,.
\eea 
The effective potential of the \thdm{} is then given by
\bea
V^{\text{eff}}_{\text{\thdm{}}}(t,\tilde t) & = & V^{\text{\lo{}}}_{\text{\thdm{}}} + V^{\text{\nlo{}}}(t) + V^{\text{\nlo{}}}({\tilde t}) + {\cal O}(\alpha^2) \,.
\eea 
The field-dependent mass parameters are defined as
\bea
\label{eq:fielddepmasses}
\overline m_t^2 & = & |X|^2 \; , \nonumber \\
\overline m_{\tilde t_{1,2}}^2 & = & \frac{1}{2} \left(  M_{\tilde t_R}^2 +
M_{\tilde t_L}^2 + 2\overline m_t^2 \mp \sqrt{( M_{\tilde t_L}^2- M_{\tilde t_R}^2)^2
+ 4|\tilde X|^2} \right)\,, \nonumber \\
X & = & h_t \phi_u^0 , \qquad
\tilde X = h_t \left[ A_t \phi_u^0 - \mu \phi_d^{0*} \right] \, .
\eea
Therein $h_t$ denotes the top-Yukawa coupling $h_t=\sqrt{2}m_t/v_u$.
If not mentioned otherwise all parameters are running parameters evaluated at the scale $Q$.
The coefficient $C_\varepsilon =\Gamma(1+\varepsilon)(4\pi)^\varepsilon$ expands
to $1+(-\gamma_E+\log(4\pi))\varepsilon$ for small $\varepsilon$ and results
in the ultraviolet divergent term $\Delta_\varepsilon=\tfrac{1}{\varepsilon}-\gamma_E+\log(4\pi)$.
The renormalization scale $Q$ is a priori not fixed but should be of
${\cal O}(M_S)$. It should be noted that the scale $Q$ represents the
matching scale between the full \mssm{} and the low-energy \thdm{}. Below this
scale the top quark and the stop states are integrated out and thus do
not contribute to the low-energy running of the Higgs self-interactions,
i.e. they are decoupled. On the other hand the low-energy self-couplings
$\lambda_{ijk}$ develop a residual scale dependence due to the light
particles still present in the low-energy \thdm{} spectrum. \s

We are interested both in the corrections to the Higgs masses and to
the triple Higgs self-couplings in order to compare differences directly.
Some of our subsequent discussion is a historical review, but allows to understand
the underlying basis of the additional terms that we add to the \hmssm{} approach.

\subsection{Corrections to the Higgs masses}
\label{sec:higgsmasscorr}
In this section we discuss the corrections to the Higgs masses obtained
from the \thdm{} effective potential.
Within the discussion we keep top-quark and stop contributions separated.
In order to obtain the \cp{}-even Higgs masses, the \thdm{} effective potential
is expanded in the \cp{}-even and \cp{}-odd components of the neutral fields $\phi_d^0$
and $\phi_u^0$, see \eqn{eq:neutralfieldsexpand}.\footnote{Our calculation is based on the field definition of the full \mssm{},
the difference to the fields of the effective low-energy \thdm{} is given by the
kinetic mixing described in \sct{sec:efflag}.} The mass corrections are then obtained as second derivatives
with respect to these components. A nice explanation of this procedure is
given in \citere{Brucherseifer:2012eol}, which reproduces the corrections
presented in the original
publications~\cite{Okada:1990vk,Haber:1990aw,Ellis:1990nz,Ellis:1991zd}.
The symmetric mass corrections $\Delta \mathcal{M}^2_{ij}$
in gauge eigenstates are added to the tree-level mass
matrix $\mathcal{M}^2_{\text{tree}}$ in \eqn{eq:treemassmssm}. We split them according to
\bea
\label{eq:corrmassmssm}
\Delta \mathcal{M}_{ij}^2=\Delta \mathcal{M}_{ij}^{2}(t)+\Delta \mathcal{M}_{ij}^{2}(\tilde t)
\eea
in top-quark and stop-induced corrections originating from $V^{\text{\nlo{}}}(t)$ and $V^{\text{\nlo{}}}(\tilde t)$, respectively.
The top-induced correction contributes only to $\Delta\mathcal{M}_{uu}^2(t)$.
The corrections are given by
\bea\nonumber
\Delta \mathcal{M}_{uu}^2(t) &=& \frac{12}{(4\pi)^2v^2s_\beta^2}m_t^4\left[2\Delta_\varepsilon +
2\log\left(\frac{Q^2}{m_t^2}\right)\right]\,, \\\nonumber
\Delta \mathcal{M}_{uu}^2(\tilde t) &=&
\frac{12}{(4\pi)^2v^2s_\beta^2}m_t^4\left[-2\Delta_\varepsilon +
A_t^2C_t^2g_t+2A_tC_t\log\left(\frac{\msto^2}{\mstt^2}\right)+2\log\left(\frac{\msto\mstt}{Q^2}\right)\right]\,,\\\nonumber
\Delta \mathcal{M}_{dd}^2(\tilde t) &=& \frac{12}{(4\pi)^2v^2s_\beta^2}m_t^4C_t^2\mu^2g_t\,,\\
\Delta \mathcal{M}_{du}^2(\tilde t) &=& -\frac{12}{(4\pi)^2v^2s_\beta^2}m_t^4
C_t\mu\left[A_tC_tg_t+\log\left(\frac{\msto^2}{\mstt^2}\right)\right]\,,
\eea
where we have used the abbreviations
\bea
\label{eq:ctabb}
C_t=\frac{X_t}{\msto^2-\mstt^2}\,,\qquad g_t=2-\frac{\msto^2+\mstt^2}{\msto^2-\mstt^2}\log\frac{\msto^2}{\mstt^2}\,.
\eea
Two remarks are in order: Only through a non-vanishing $\mu$-term
also the corrections $\Delta \mathcal{M}_{dd}^2$ and $\Delta \mathcal{M}_{du}^2$ are non-zero.
The ultraviolet divergent terms and the renormalization scale dependence cancel in the sum of top-quark and stop-induced corrections:
\bea
\label{eq:DeltaMuu2}
\Delta \mathcal{M}_{uu}^2 = \frac{12}{(4\pi)^2v^2s_\beta^2}m_t^4\left[A_t^2C_t^2g_t+2A_tC_t\log\left(\frac{\msto^2}{\mstt^2}\right)+2\log\left(\frac{\msto\mstt}{m_t^2}\right)\right]\,.
\eea
It is therefore obvious that an effective \thdm{}, where both the top quark and stops
are taken into account in the effective potential, yields finite
corrections at NLO. In contrast in an effective \thdm{}, which only
involves tops, also ultraviolet divergences appear, which can only be cured through a proper
renormalization. We will discuss this when we will later calculate our partial decay width.

\subsection{The $\epsilon$ approximation for the light Higgs mass}
Our previous discussion of the loop corrections was carried out in gauge eigenstates.
We can rotate to mass eigenstates to obtain the correction
to the light Higgs mass, $\Delta M_h$. We now consider the case where
left- and right-handed soft-\susy{} breaking mass parameters $M_S=M_{\tilde t_L}=M_{\tilde t_R}$
are identical. We expand the mass corrections in inverse powers of $M_S$ and
subsequently present the result for $\Delta \mathcal{M}_{uu}^2$ and the total mass correction~$\Delta M_h^2$:
\bea\nonumber
\Delta \mathcal{M}_{uu}^2 &=& \frac{3 G_F}{\sqrt{2}\pi^2s^2_\beta}m_t^4
\left[ \log \left(\frac{M_S^2}{m_t^2}\right) +
  \frac{X_tA_t}{M_S^2} \left( 1- \frac{X_tA_t}{12M_S^2} \right) \right] \,, \\
\Delta M_h^2 &=& \frac{3 G_F}{\sqrt{2}\pi^2}m_t^4
\left[ \log \left(\frac{M_S^2}{m_t^2}\right) +
  \frac{X_t^2}{M_S^2} \left( 1- \frac{X_t^2}{12M_S^2} \right) \right] \,.
\label{eq:epsilondefinition}
\eea
Therein, we expressed the \sm{} VEV through the Fermi constant $G_F$.
Moreover for $\Delta M_h^2$ we employed the decoupling limit $\alpha\to \beta-\pi/2$,
which is associated with $M_A\gg M_Z$.
The last well-known relation in \eqn{eq:epsilondefinition}, see \citeres{Okada:1990vk,Haber:1990aw,Ellis:1990nz,Ellis:1991zd}
and later updates in \citeres{Carena:1995bx,Haber:1996fp}, shows that the light
Higgs mass grows with the fourth power of the top-quark mass and
logarithmically with the stop masses associated with the \susy{} scale $M_S$.
We define the parameter $\epsilon = \Delta\mathcal{M}_{uu}^2$. Setting $\mu=0$
and thus $X_t=A_t$ we see that $\epsilon$ exactly corresponds to the correction $\Delta M_h^2/s_\beta^2$
in accordance with the fact that all other elements of $\Delta \mathcal{M}_{ij}^2$ in gauge eigenstates vanish.
The $\epsilon$ correction increases the upper mass bound to
\beq
M_h^2 \le M_Z^2 \cos^2 2\beta + \epsilon \sin^2 \beta \,.
\eeq
The mass of the lightest Higgs boson is explicitly given by
\bea
\label{eq:LHmassepsilonapprox}
M_h^2 &=& \frac{1}{2} \left[ M_A^2 + M_Z^2 + \epsilon  - \sqrt{ (M_A^2 + M_Z^2 + \epsilon)^2 - 4 M_A^2 M_Z^2
    c^2_{2\beta}- 4 \epsilon (M_A^2 s^2_\beta + M_Z^2 c^2_\beta) }  \right] \,.
\eea
Within this approximation
the masses of the heavy neutral and charged
Higgs bosons are obtained by sum rules,
\bea
\label{eq:HHmassepsilonapprox}
M_H^2 &=& M_A^2 + M_Z^2 - M_h^2 + \epsilon \,,\nonumber \\
M_{H^\pm}^2 &=& M_A^2 + M_W^2 \,,
\eea
and the effective mixing parameter $\alpha$ between the \cp{}-even scalars is
given by 
\beq
\label{eq:alphaepsilonapprox}
\tan 2 \alpha = \tan 2 \beta \frac{M_A^2 + M_Z^2}{M_A^2 - M_Z^2 +
\epsilon/\cos 2\beta}\,.
\eeq
We will later come back to the $\epsilon$ approximation also for the
Higgs self-couplings, but before introduce the \hmssm{} approach.

\subsection{The hMSSM approach}
\label{sec:hMSSMapproach}
As argued before, in the limit of a $\mu$-term, which is small compared to the stop spectrum,
the corrections $\epsilon$ and $\Delta M_h^2/s_\beta^2$
in \eqn{eq:epsilondefinition} are identical. Then, the complete correction to the light
Higgs mass originating from the top-quark and stop sector only enters the element $\Delta \mathcal{M}_{uu}^2$.
The \hmssm{} approach~\cite{Maiani:2013hud,Djouadi:2013vqa,Djouadi:2013uqa,Djouadi:2015jea}
assumes all supersymmetric particles to be heavy and 
is agnostic for what concerns the origin of the mass corrections. Instead in the \hmssm{} one
obtains $\Delta \mathcal{M}_{uu}^2$ by inverting \eqn{eq:LHmassepsilonapprox} using $M_h$ as an input parameter, which yields
\bea
\label{eq:hMSSMapproach}
\epsilon = \Delta \mathcal{M}_{uu}^2 = \frac{M_h^2(M_A^2+M_Z^2-M_h^2)-M_A^2M_Z^2c^2_{2\beta}}{M_Z^2c_\beta^2+M_A^2s_\beta^2-M_h^2} \,.
\eea 
The heavy Higgs mass $M_H$ and the mixing angle $\alpha$ are then in turn also fixed to the values
in \eqn{eq:HHmassepsilonapprox} and \eqn{eq:alphaepsilonapprox}, respectively.
The limitations of this procedure are rather obvious: 
First, the corrections of the top-quark and stop sector are assumed to be dominant, which is true at low values of $\tan\beta$,
while at larger values of $\tan\beta$ also corrections from the bottom
and sbottom sector provide a potentially relevant (subleading) contribution.
Second, neglecting the corrections $\Delta \mathcal{M}_{dd}^2$ and $\Delta \mathcal{M}_{du}^2$
is only compatible with a $\mu$-term, which is small compared to the stop spectrum and thus implies relatively light Higgsinos.
This means not all supersymmetric particles are necessarily heavy. They can influence Higgs physics mainly
through decays, either by allowing for additional decay channels for heavy Higgs bosons or through
contributions to loop-induced decays of the Higgs bosons.

\subsection{The improved hMSSM for Higgs self-couplings} \label{sec:improve}
We are interested in the corrections to the triple Higgs self-couplings
presented in \eqn{eq:LOlambdas} that emerge from the effective potential in \eqn{eq:V1mssm}. Corrections
are obtained by performing the third derivatives with respect to the
corresponding Higgs fields.
Again, we only focus on corrections from the top-quark and stop sector and thus split the individual contributions as follows
\bea
\Delta\lambda_{ijk} = \Delta\lambda_{ijk}(t)+\Delta\lambda_{ijk}(\tilde t)\,,
\eea
both in gauge eigenstates $\{ i,j,k\} \in \{ d,u\}$
and in mass eigenstates $\{ i,j,k\} \in \{ h,H\}$.
We present them subsequently 
in gauge eigenstates. 
The top-quark contribution enters $\Delta\lambda_{uuu}$, while for
a non-vanishing $\mu$-term the stop contributions yield a correction
to all couplings in accordance with the original results in \citere{Barger:1991ed}:
{\allowdisplaybreaks
\bea \nonumber
\Delta\lambda_{uuu}(t)&=&\frac{72}{(4\pi)^2v^3s_\beta^3}m_t^4\left[\Delta_\varepsilon-\frac{2}{3}+\log\left(\frac{Q^2}{m_t^2}\right)\right]\,,\\\nonumber
\Delta\lambda_{uuu}(\tilde t)&=&
-\frac{72}{(4\pi)^2v^3s_\beta^3}m_t^4\left[\Delta_\varepsilon+\log\left(\frac{Q^2}{\mstt^2}\right)\right]
+\frac{12}{(4\pi)^2v^3s_\beta^3}m_t^4 \times\\\nonumber
&&\times \left\{2A_t^3C_t^3m_t^2\left[6\frac{g_t}{\mstt^2-\msto^2}+\frac{\mstt^2-\msto^2}{\msto^2\mstt^2}\right]-3\frac{A_t^3C_tg_t}{\mstt^2-\msto^2}+3\frac{A_t^2(2-g_t)}{\msto^2+\mstt^2}\right.\\\nonumber
&&+6\frac{A_t^2C_t^2m_t^2}{\msto^2+\mstt^2}\left[2g_t+\frac{(\mstt^2-\msto^2)^2}{\msto^2\mstt^2}\right]-3\frac{\mstt^2-\msto^2}{\msto^2+\mstt^2}(2-g_t)\\\nonumber
&&\left.-3A_tC_t(\mstt^2-\msto^2)\left[\frac{2-g_t}{\msto^2+\mstt^2}-2\frac{m_t^2}{\msto^2\mstt^2}\right]+2m_t^2\frac{\msto^2+\mstt^2}{\msto^2\mstt^2}\right\}\,,\\\nonumber
\Delta\lambda_{ddd}(\tilde t)&=&-\frac{12}{(4\pi)^2v^3s_\beta^3}m_t^4C_t\mu^3(\mstt^2-\msto^2)\left\{2\frac{C_t^2m_t^2}{\msto^2\mstt^2}+3\frac{4C_t^2m_t^2-1}{(\mstt^2-\msto^2)^2}g_t\right\}\,,\\\nonumber
\Delta\lambda_{ddu}(\tilde t)&=&\frac{12}{(4\pi)^2v^3s_\beta^3}m_t^4\mu^2\left\{2A_tC_t^3m_t^2\left[6\frac{g_t}{\mstt^2-\msto^2}+\frac{\mstt^2-\msto^2}{\msto^2\mstt^2}\right]\right.\\\nonumber
&&\left.-3\frac{A_tC_tg_t}{\mstt^2-\msto^2}+2\frac{C_t^2m_t^2}{\msto^2+\mstt^2}\left[2g_t+\frac{(\mstt^2-\msto^2)^2}{\msto^2\mstt^2}\right]+\frac{2-g_t}{\msto^2+\mstt^2}\right\}\,,\\\nonumber
\Delta\lambda_{duu}(\tilde t)&=&-\frac{12}{(4\pi)^2v^3s_\beta^3}m_t^4\mu\left\{2A_t^2C_t^3m_t^2\left[6\frac{g_t}{\mstt^2-\msto^2}+\frac{\mstt^2-\msto^2}{\msto^2\mstt^2}\right]\right.\\\nonumber
&&-3A_t^2C_t\frac{g_t}{\mstt^2-\msto^2}+4A_tC_t^2m_t^2\frac{\mstt^2-\msto^2}{\msto^2+\mstt^2}\left[2\frac{g_t}{\mstt^2-\msto^2}+\frac{\mstt^2-\msto^2}{\msto^2\mstt^2}\right]\\
&&\left.+2A_t\frac{2-g_t}{\msto^2+\mstt^2}+C_t(\mstt^2-\msto^2)\left[2\frac{m_t^2}{\msto^2\mstt^2}-\frac{2-g_t}{\msto^2+\mstt^2}\right]\right\}\,.
\eea}
The ultraviolet divergences, which are only part of $\Delta\lambda_{uuu}$,
cancel between the top-quark and stop contributions $\Delta\lambda_{uuu}(t)$
and $\Delta\lambda_{uuu}(\tilde t)$, respectively. So does the
renormalization scale dependence related to the top-quark and stop
contributions.
It is apparent that in contrast to the calculation of the Higgs mass terms the top-quark
induced correction develops an additional constant factor~$(-\tfrac{2}{3})$ at the one-loop level. 
\s

Since we are interested in the decay $H\to hh$
we combine the results above to obtain $\Delta\lambda_{Hhh}$ by
rotating with the \cp{}-even mixing angle $\alpha$. This yields
{\allowdisplaybreaks
\bea \nonumber
\Delta \lambda_{Hhh}&=&-s_\alpha c_\alpha^2\Delta\lambda_{uuu}
-c_\alpha(c_\alpha^2-2s_\alpha^2)\Delta \lambda_{duu}
+s_\alpha(2c_\alpha^2-s_\alpha^2)\Delta \lambda_{ddu}
-s_\alpha^2c_\alpha \Delta \lambda_{ddd}\\\nonumber
&=&\frac{12}{(4\pi)^2v^3s_\beta^3}m_t^4s_\alpha c_\alpha^2\left\{-4+6\log\left(\frac{\msto\mstt}{m_t^2}\right)+(\msto^2-\mstt^2)C_t(E_t+2F_t)\log\left(\frac{\msto^2}{\mstt^2}\right)\right.\\\nonumber
&&+F_t(\msto^2-\mstt^2)(1-4m_t^2C_t^2)\left[3C_tE_tF_tg_t(\msto^2-\mstt^2)+(2E_t+F_t)\log\left(\frac{\msto^2}{\mstt^2}\right)\right]\\\nonumber
&&+2\left[\frac{m_t^2}{\msto^2}\left[1+(\msto^2-\mstt^2)C_tE_t\right]\left[1+(\msto^2-\mstt^2)C_tF_t\right]^2\right]\\
&&+\left.2\left[\frac{m_t^2}{\mstt^2}\left[1-(\msto^2-\mstt^2)C_tE_t\right]\left[1-(\msto^2-\mstt^2)C_tF_t\right]^2\right]\right\}\,.
\label{eq:deltalambdaHhhfull}
\eea}
Therein we introduced the additional abbreviations
\bea
E_t &=&\frac{A_t-\mu\cot\alpha}{\msto^2-\mstt^2},\qquad F_t=\frac{A_t+\mu\tan\alpha}{\msto^2-\mstt^2}\,.
\eea
For degenerate soft-\susy{} breaking masses $M_S=M_{\tilde t_L}=M_{\tilde t_R}$ and
in the limit $\mu \ll M_S$ the expansion in inverse powers of the \susy{} scale $M_S$ implies
\bea\notag
\Delta \lambda_{Hhh}&=&\frac{72s_\alpha c_\alpha^2}{(4\pi)^2v^3s_\beta^3}m_t^4\left [\log\left(\frac{M_S^2}{m_t^2}\right)
+\frac{A_t^2}{M_S^2}\left(1-\frac{A_t^2}{12 M_S^2}\right)\right.\\
&&\left.-\frac{2}{3}+\frac{5m_t^2}{3M_S^2}-\frac{5A_t^2 m_t^2}{2M_S^4}+\frac{5A_t^4m_t^2}{6M_S^6}-\frac{A_t^6m_t^2}{12M_S^8}
\right]\,.
\label{eq:deltalambdaHhhwoeps}
\eea
Here our expansion in inverse powers of $M_S$ adds terms proportional to $m_t^6$
and all relevant terms that are of order $1/M_S^2$ for $A_t\sim M_S$.
If we perform a similar expansion in $\mathcal{M}_{uu}^2$ in \eqn{eq:DeltaMuu2}, we obtain
\bea
\label{eq:epsilondefinition2}
\epsilon = \frac{3 G_F}{\sqrt{2}\pi^2s^2_\beta}m_t^4
\left[ \log \left(\frac{M_S^2}{m_t^2}\right) +
  \frac{A_t^2}{M_S^2} \left( 1- \frac{A_t^2}{12M_S^2} \right)  + \frac{m_t^2}{M_S^2} - \frac{3m_t^2A_t^2}{2M_S^4} + \frac{m_t^2A_t^4}{2M_S^6} - \frac{m_t^2A_t^6}{20M_S^8}\right]\,.
\eea
In $\Delta \lambda_{Hhh}$ we can identify the corrections that are part of $\epsilon$ in \eqn{eq:epsilondefinition2}, and therefore write
\bea
\label{eq:deltalambdaHhh}
\Delta \lambda_{Hhh}=\frac{3s_\alpha c_\alpha^2}{vs_\beta}\left[\epsilon+\frac{24}{(4\pi)^2v^2s_\beta^2}
m_t^4\left(-\frac{2}{3}+\frac{2m_t^2}{3M_S^2}-\frac{m_t^2A_t^2}{M_S^4}+\frac{m_t^2A_t^4}{3M_S^6}-\frac{m_t^2A_t^6}{30M_S^8}\right)\right]\,.
\eea
The \hmssm{} approach advocates to just use the first term $\epsilon$ in \eqn{eq:deltalambdaHhh}
as a correction not only for the Higgs masses, but also for the Higgs self-couplings and
thus misses the second bracket. However, the second bracket includes a
purely top-induced contribution,
which originates from the top-quark correction to $\Delta \lambda_{uuu}(t)$ and should not be missed in the Higgs self-couplings,
neither in the $\epsilon$ approximation nor in the \hmssm{} approach.
The other terms of the second bracket are instead well suppressed for heavy squark masses
and in the spirit of the \hmssm{} approach can be neglected.
This constant correction beyond the terms comprised in the $\epsilon$ approximation
also appears in all other Higgs self-couplings. We therefore define
\bea
\label{eq:barepsilon}
\overline\epsilon=\epsilon - \frac{24 m_t^4}{(4\pi)^2v^2s_\beta^2}\frac{2}{3}
\eea
and obtain effective couplings of the form
\begin{align}
\lambda_{hhh}^{\overline\epsilon}  &=  \lambda_{hhh} + \frac{3c_\alpha^3}{vs_\beta} \overline\epsilon\,,\qquad
&\lambda_{Hhh}^{\overline\epsilon}  &=  \lambda_{Hhh} + \frac{3s_\alpha c_\alpha^2}{vs_\beta} \overline\epsilon\,,
\nonumber \\
\lambda_{HHh}^{\overline\epsilon}  &=  \lambda_{HHh} + \frac{3 s_\alpha^2c_\alpha}{vs_\beta}\overline\epsilon\,,\qquad
&\lambda_{HHH}^{\overline\epsilon}  &=  \lambda_{HHH} +
\frac{3s_\alpha^3}{vs_\beta}\overline\epsilon\,,
\nonumber \\
\lambda_{hAA}^{\overline\epsilon}  &=  \lambda_{hAA} + \frac{c_\alpha c_\beta^2}{vs_\beta}\overline\epsilon\,,\qquad
&\lambda_{HAA}^{\overline\epsilon}  &=  \lambda_{HAA} + \frac{s_\alpha c_\beta^2}{vs_\beta}\overline\epsilon\,,
\label{eq:efflambdas}
\end{align}
where the tree-level couplings are taken from \eqn{eq:LOlambdas}.
The usage of the effective couplings $\lambda_{ijk}^{\overline\epsilon}$
can be considered an improvement of the original \hmssm{} approach, which is why we dub it ``improved \hmssm{}''.
Additionally, we will later use the couplings~$\lambda_{ijk}^\epsilon$,
which are defined as in \eqn{eq:efflambdas}, but with $\epsilon$ instead of $\overline\epsilon$.
Thus, they correspond to the original \hmssm{} approach. 
Note that $\epsilon$ and therefore also $\overline\epsilon$ can be calculated either from the actual
correction $\epsilon=\Delta \mathcal{M}_{uu}^2$, which equals the $\epsilon$
approximation, or according to the \hmssm{} approach.
We follow the \hmssm{} approach and thus obtain $\epsilon$
from the right-hand side of \eqn{eq:hMSSMapproach} and
$\overline\epsilon$ from \eqn{eq:barepsilon}. \s

We summarize that in our subsequent calculation of the partial decay width $H\to hh$ we 
work with effective low-energy \thdm{} couplings and mixing angles
and thus do not only employ the tree-level coupling $\lambda_{Hhh}$
in the leading-order amplitude. We instead use the effective couplings
$\lambda_{Hhh}^{\epsilon}$ and $\lambda_{Hhh}^{\overline\epsilon}$ of
the original and the improved \hmssm{} approach, respectively,
and, moreover, make use of the complete correction in the effective potential given by
\bea
 \lambda_{Hhh}^{\text{eff}} (t,\tilde t) = \lambda_{Hhh} + \Delta
\lambda_{Hhh}(t,\tilde t) \quad [\text{\eqn{eq:deltalambdaHhhfull}}]
\eea
that corresponds to the proper matching of the low-energy
\thdm{} to the \mssm{}.
We add the arguments $(t,\tilde t)$ to $\lambda_{Hhh}^{\text{eff}}$, which have to be understood as flags,
i.e. we allow to add the top- and stop-induced contribution to $\Delta \lambda_{Hhh}$ separately.
In practice at \lo{} in the decay width we will always use $\lambda_{Hhh}^{\text{eff}} (1,1)$,
which is ultraviolet finite in contrast to $\lambda_{Hhh}^{\text{eff}} (1,0)$.

\section{The partial decay width $H\to hh$} \label{sec:calc}

For now we have discussed both the \mssm{} and its approximation, the \hmssm{},
in the effective potential approach, which allows us to match
them to an effective \thdm{}. In our subsequent discussion the couplings and masses of the \thdm{}
include the previously mentioned one-loop corrections.
We explicitly provided formulas for the corrected Higgs self-couplings
and note that the Higgs masses are obtained by diagonalizing the
one-loop corrected mass matrices in the \epa{},
see also \citere{Brucherseifer:2013qva}.
For the scope of this work we neglect the contributions of other particles than the top
quark and squarks that, however, could be taken into account in a full
diagrammatic calculation in a straightforward manner. In the same
context we will also neglect the residual RGE-evolution of the Higgs
self-couplings $\lambda_{ijk}$ within the low-energy \thdm{} in the
following, i.e.~work with the values obtained at the matching scale. In
a full calculation the running due to the light degrees of freedom would
have to be taken into account for consistency. Since the dominant radiative
corrections to the decay $H\to hh$ are known to emerge from top-quark and stop
loops, these neglected effects are only subleading and do not contribute to the
mismatch between the full \mssm{} and the \thdm{} as the low-energy limit. \s

\begin{figure}
  \begin{center}
    \begin{tabular}{ccccc}
      \includegraphics[width=0.17\textwidth]{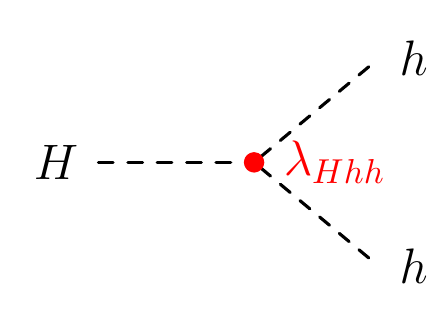} &
      \includegraphics[width=0.17\textwidth]{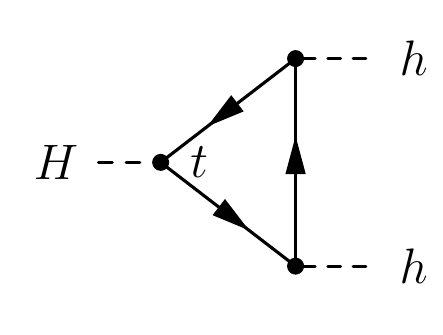} &
      \includegraphics[width=0.17\textwidth]{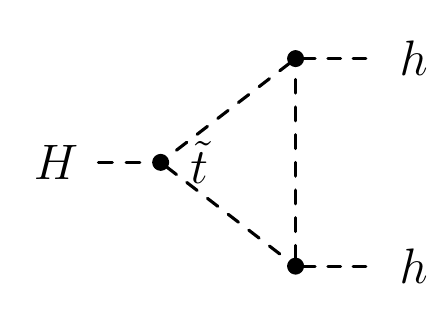} &
      \includegraphics[width=0.17\textwidth]{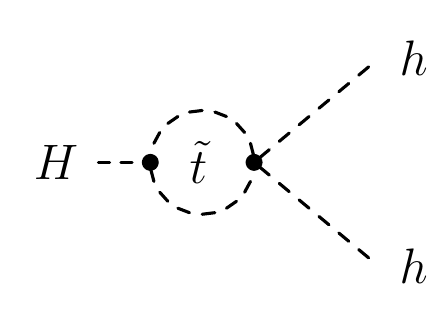} &
      \includegraphics[width=0.17\textwidth]{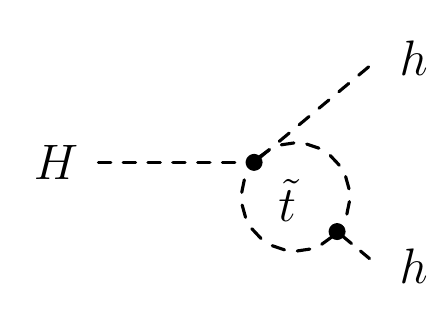}
      \\[-0.5cm]
      (a) & (b) & (c) & (d) & (e)\\
      \includegraphics[width=0.17\textwidth]{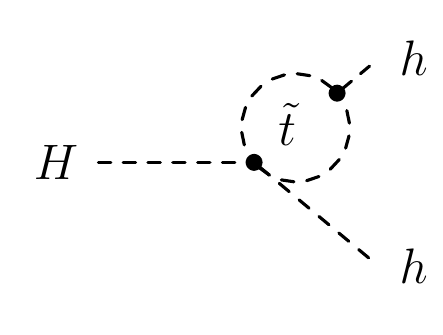} &
      \includegraphics[width=0.17\textwidth]{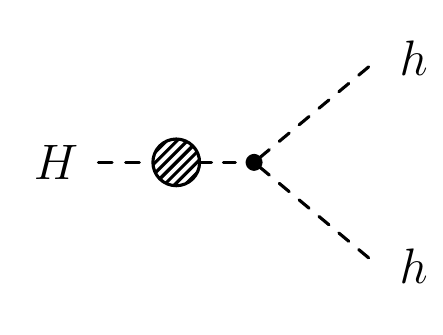} &
      \includegraphics[width=0.17\textwidth]{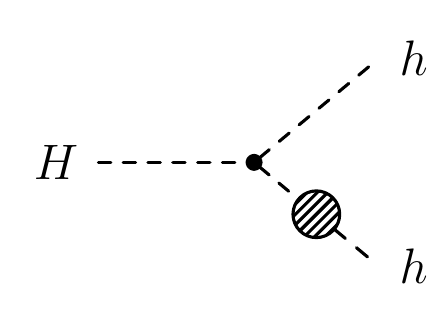} &
      \includegraphics[width=0.17\textwidth]{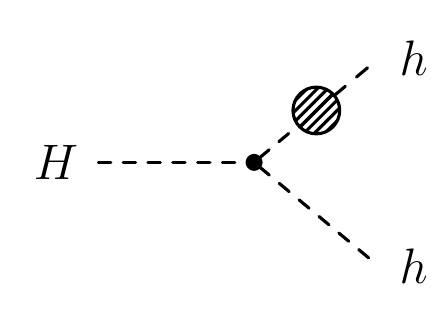} &
      \includegraphics[width=0.17\textwidth]{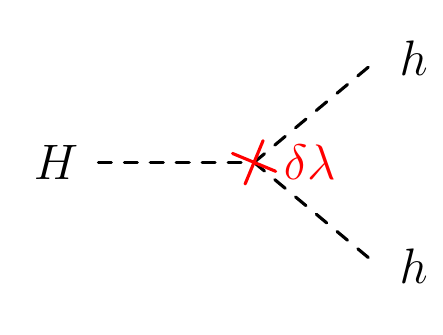} \\[-0.5cm]
      (f) & (g) & (h) & (i) & (j)\\
      \parbox{0.16\textwidth}{\includegraphics[width=0.17\textwidth]{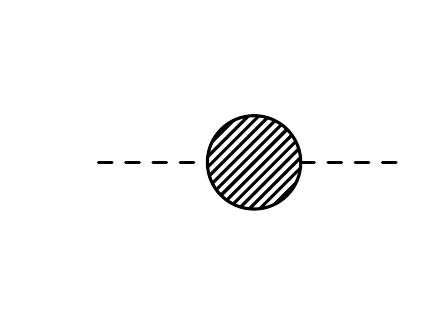}}$=$ &
      \parbox{0.16\textwidth}{\includegraphics[width=0.17\textwidth]{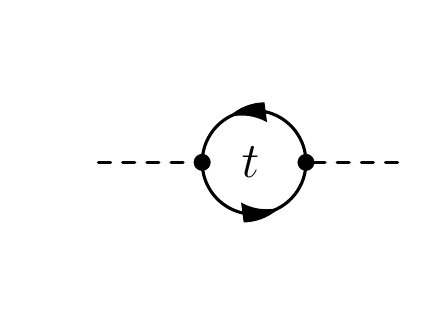}} &
      \parbox{0.16\textwidth}{\includegraphics[width=0.17\textwidth]{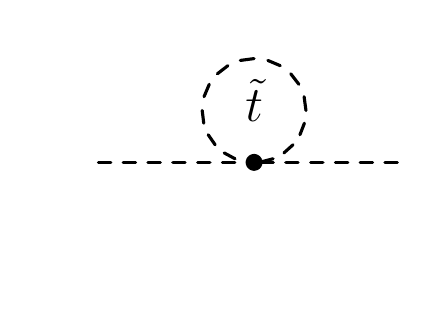}} &
      \parbox{0.16\textwidth}{\includegraphics[width=0.17\textwidth]{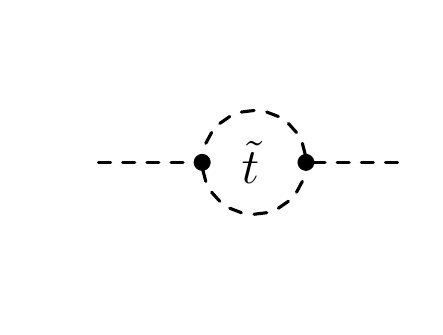}} &
      \parbox{0.16\textwidth}{\includegraphics[width=0.17\textwidth]{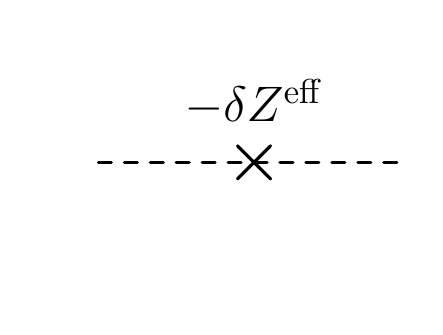}}\\[-0.5cm]
      (k) & (l) & (m) & (n) & (o)
    \end{tabular}
    \parbox{\textwidth}{
      \caption[]{\label{fig:feynman} Feynman diagrams for $H\to hh$: (a) Tree-level diagram; (b)--(f) virtual one-loop corrections;
      (g)--(j) external self-energy corrections and counterterm; (k) generic self-energy that comprises the self-energy corrections 
      and the kinetic counterterm depicted in (l)--(o).}}
  \end{center}
\end{figure}

In this section we calculate the partial decay width $H\to hh$.
We perform a Feynman-diagrammatic calculation at the one-loop level including
the full momentum-dependent corrections.
We denote the momentum of the
incoming Higgs boson by $q_H$ and the momenta
of the outgoing Higgs bosons by $q_1$ and $q_2$. Ultimately, we perform
an on-shell calculation and thus set $q_H^2=M_H^2$ and $q_1^2=q_2^2=M_h^2$.
In the Feynman-diagrammatic approach we obtain the partial decay
width $\Gamma(H\to hh)$ according to
\bea
\Gamma(H\to hh)=\frac{|\mathcal{A}|^2}{32\pi M_H}\sqrt{1-\frac{4M_h^2}{M_H^2}}\,,
\eea
where $\mathcal{A}=\mathcal{A}^{\text{\lo{}}}+\mathcal{A}^{\text{\nlo{}}}$ denotes the amplitude.
At pure tree-level the contribution $\mathcal{A}^{\text{\lo{}}}$ equals the Higgs self-coupling
$\lambda_{Hhh}$ as given in \eqn{eq:LOlambdas}.
The corresponding Feynman diagram is shown in \fig{fig:feynman}~(a).
However, we want to work in the effective \thdm{}, i.e. we employ
the effective couplings $\lambda_{Hhh}^{\text{eff}}$ and $\lambda_{Hhh}^{\overline\epsilon}$ and thus
incorporate higher-order effects already in $\mathcal{A}^{\text{\lo{}}}$. 
We also apply the previously discussed one-loop corrections in the \epa{},
see also \citere{Brucherseifer:2013qva}, to obtain
the external Higgs masses. \s

The loop-corrected amplitude $\mathcal{A}^{\text{\nlo{}}}$ can be split into the following
pieces
\bea
\mathcal{A}^{\text{\nlo{}}}(t,\tilde t)
=\mathcal{A}^{\text{virt}}(t,\tilde t)
+\mathcal{A}^{\text{ext}}(t,\tilde t)
+\mathcal{A}^{\text{ext,eff}}(t,\tilde t)
+\mathcal{A}^{\delta\lambda}(t,\tilde t)\,,
\eea
where $\mathcal{A}^{\text{virt}}(t,\tilde t)$ denotes the momentum-dependent
one-particle irreducible Feynman diagrams. They are depicted in \fig{fig:feynman}~(b)-(f).
$\mathcal{A}^{\text{ext}}(t,\tilde t)$ are external self-energy
corrections adjusted to the amputated Green's functions including a mixing between $H$ and $h$, see \fig{fig:feynman}~(g)-(i).
The contribution $\mathcal{A}^{\text{ext,eff}}(t,\tilde t)$ originates
from the kinetic mixing $Z^{\text{eff}}$ already discussed in
\sct{sec:efflag} and thus provides the proper normalization of the
(effective) Higgs fields in the effective low-energy \thdm{}.
Finally $\mathcal{A}^{\delta\lambda}(t,\tilde t)$ comprises additional
counterterms induced by the effective couplings of the \epa{} and is generically depicted in \fig{fig:feynman}~(j).
$\mathcal{A}^{\text{virt}}$ can be easily expressed in terms of
Passarino-Veltman integrals \cite{tHooft:1978jhc, Passarino:1978jh}. 
We present the corresponding analytic expression in \appref{app:analyticresults}.
On the other hand we have to define a renormalization scheme to fix all the remaining counterterms.
By adding the arguments $(t,\tilde t)$, which have to be understood as a flag to include or to disregard top-quark and stop contributions,
we emphasize that we can add both contributions separately to all individual ingredients of the one-loop amplitude,
see \appref{app:analyticresults} for an explanation.

\subsection{Self-energy corrections and renormalization}
\label{sec:wvandrenorm}
We employ effective couplings and effective masses in an effective \thdm{} at tree-level
and also use them in our one-loop Feynman-diagrammatic calculation.
This implies that beyond the corrections already implemented through the effective
potential, we add only momentum-dependent terms. First we discuss 
the external self-energies, which boil down to the following contributions
\vspace{-3mm}
\bea
\delta Z_H=\Sigma'_{HH}(M_H^2)\,, \quad
\delta Z_h=\Sigma'_{hh}(M_h^2)\,,\quad 
\delta Z_{Hh}(p^2)=\frac{\Sigma_{Hh}(p^2)}{M_H^2-M_h^2}\,,
\eea
where $\Sigma_{ij}(p^2)$ is the self-energy involving the two mass eigenstates
$i,j\in \{h,H\}$ and $\Sigma'_{ij}(p^2)$ is its derivative with respect to the squared external
momentum $p^2$. The self-energy corrections $\delta Z$ all enter
$\mathcal{A}^{\text{ext}}$.
Again these corrections can be expressed in terms of
Passarino-Veltman integrals and can be split into top-quark and
stop-induced corrections, see \appref{app:analyticresults}.
The self-energy corrections enter the amplitude as follows:
\bea
\label{eq:Aext}
\mathcal{A}^{\text{ext}}(t,\tilde t) = \lambda_{Hhh}(\tfrac{1}{2}\delta Z_H
+\delta Z_h)+\lambda_{hhh}\delta Z_{Hh}(M_H^2)-2\lambda_{HHh}\delta Z_{Hh}(M_h^2) 
\eea
Additionally we have to take into account the kinetic corrections explained
in \sct{sec:efflag}. They imply an additional contribution at the one-loop level
\bea
\label{eq:Aexteff}
\mathcal{A}^{\text{ext,eff}}(t,\tilde t) =
\lambda_{Hhh}(-\tfrac{1}{2}\delta Z^{\text{eff}}_H-\delta
Z^{\text{eff}}_h)-\lambda_{hhh}\delta
Z^{\text{eff}}_{Hh}(M_H^2)+2\lambda_{HHh}\delta
Z^{\text{eff}}_{Hh}(M_h^2)
\eea
that corresponds to the matching of the kinetic term of the full theory
to the kinetic term of the effective low-energy \thdm{} after integrating
out the top-quark (stops). Since our calculation is based on the field
definition of the full \mssm{} we have to divide through $Z^{\text{eff}}$
to obtain the field normalization in the effective low-energy \thdm{}, which
explains the subtraction of the corresponding terms $\delta Z^{\text{eff}}$.
The effective $Z^{\text{eff}}$ matrix is given by
\bea
\delta Z^{\text{eff}}_H = \Sigma'_{HH}(0)\,, \qquad \delta Z^{\text{eff}}_h 
= \Sigma'_{hh}(0)\,,\qquad \delta Z^{\text{eff}}_{Hh}(p^2)=\frac{p^2\Sigma'_{Hh}(0)}{M_H^2-M_h^2}\,.
\eea
Also for the self-energies presented here, we can take into account top-quark
and stop contributions separately. \s

The mixing angle $\alpha$ is renormalized by promoting the tree-level relation
for $\alpha_0=\alpha+\delta\alpha$, see \eqn{eq:treealpha},
to the one-loop level, which results in
\bea
\delta \alpha = -\frac{s_{4\alpha}}{4}\left(\frac{\Delta \mathcal{M}_{du}^2}{\mathcal{M}_{du}^2}
-\frac{\Delta \mathcal{M}_{uu}^2-\Delta \mathcal{M}_{dd}^2}{\mathcal{M}_{uu}^2- \mathcal{M}_{dd}^2}\right)\,.
\eea
Using the explicit expressions for the mass corrections in the \epa{}, see \eqn{eq:corrmassmssm},
implies that the mixing angle
is renormalized for vanishing external momenta in accordance with the
consistent definition of effective low-energy parameters. \s

The tree-level coupling $\lambda_{Hhh}$ in \eqn{eq:LOlambdas} suggests
that we need to renormalize not only the mixing angle $\alpha$, but also
the angle $\beta$ as well as the electroweak sector, i.e. $M_Z$ and $v$.
However, since we are working with effective parameters defined by the
radiatively corrected \epa{} in our one-loop corrections, the
renormalization of the parameters $\beta, M_Z$ and $v$ is already part
of the renormalization of the effective potential intrinsically, since
the full counterterm $\delta \lambda_{Hhh}$ is defined by the \epa{} that
determines the full mismatch between the \mssm{} and the low-energy \thdm{}.
This is different for the mixing angle $\alpha$ that enters as the
external rotation of the current eigenstates to the mass eigenstates
applied to the fully corrected and renormalized effective potential
in the current-eigenstate basis. We can therefore write
\bea
\mathcal{A}^{\delta\lambda}(t,\tilde t) = \frac{\partial
\lambda_{Hhh}}{\partial \alpha} \delta\alpha + \mathcal{A}^{\text{eff}}
= \lambda_{hhh} \delta\alpha - 2\lambda_{HHh}\delta \alpha +
\mathcal{A}^{\text{eff}}\,.
\eea
The renormalization of the mixing angle $\alpha$ is formally part of
$\mathcal{A}^{\delta \lambda}$, but adds to the non-diagonal
renormalization factor $\delta Z_{Hh}$, which is obvious from the
relation
$\tfrac{\partial\lambda_{Hhh}}{\partial\alpha}=\lambda_{hhh}-2\lambda_{HHh}$.
Since we employ an effective coupling in the tree-level amplitude
$\mathcal{A}^{\text{\lo{}}}$, we need to adjust the
counterterm~$\mathcal{A}^{\text{eff}}$ accordingly to avoid double-counting. We
obtain
\bea
\mathcal{A}^{\text{eff}}(t,\tilde t) = - \Delta
\lambda_{Hhh}(t,\tilde t) = -
\left.\mathcal{A}^{\text{virt}}(t,\tilde t)\right|_{q_i^2=0}\,.
\label{eq:vertexle}
\eea
We explicitly checked the last relation, which is in accordance with
\citere{Brignole:1992zv}. Again all these relations hold for top-quark and
stop contributions separately. 
It is obvious that the combination of the self-energy corrections $\mathcal{A}^{\text{ext}}$
and $\mathcal{A}^{\text{ext,eff}}$ and the counterterm $\mathcal{A}^{\delta\lambda}$
only leaves momentum-dependent corrections in the amplitude $\mathcal{A}^{\text{\nlo{}}}$.

\subsection{Combining the results}
\label{sec:combresult}
We have now presented all relevant ingredients for the calculation of
the partial decay width $H\to hh$ at the one-loop level within the
effective \thdm{}.  In the following we will present our numerical
results for different combinations of the effective or tree-level
couplings and including top-quark and squark contributions separately.
We emphasize that also just taking the top-quark contribution into
account leads to an ultraviolet finite result in
accordance with the Appelquist-Carazzone decoupling theorem
\cite{Appelquist:1974tg}. According to
Eq.~(\ref{eq:vertexle}) the contribution
$\mathcal{A}^{\text{eff}}(t,\tilde t)$ cancels all divergences of the
vertex corrections of Fig.~\ref{fig:feynman} (b)-(f). The contributions
of the counterterm $\delta\alpha$ and the counterterms $\delta
Z_{ij}^{\text{eff}}$ together cancel the divergences of the external
self-energies of Fig.~\ref{fig:feynman} (g)-(i). All these cancellations
emerge for the top-quark and stop contributions separately. On the other hand
the sum of the top-quark and stop contributions leads to finite vertex
corrections already before renormalization and in the same way to a
finite counterterm~$\mathcal{A}^{\text{eff}}$.\s

We summarize the options for our numerical analysis
in \tab{tab:widthoptions}.
Option 1 is the \mssm{} calculation involving
top-quark (and stop) loops without the use of the effective Higgs self-coupling
$\lambda_{Hhh}^{\text{eff}}$, but employing the tree-level values at \lo{} and \nlo{}.
We name the corresponding decay widths~$\Gamma^{\text{\lo{}}}$ and $\Gamma^{\text{\nlo{}}}(t,\tilde t)$.
Through the arguments $(t,\tilde t)$ we indicate whether only top quarks $(1,0)$
or additionally stops~$(1,1)$ are included in the Feynman-diagrammatic
calculation at the one-loop level.
Option 2 is the same calculation using the effective Higgs self-couplings
$\lambda_{Hhh}^{\epsilon}$ and $\lambda_{Hhh}^{\overline\epsilon}$
within the original and our improved \hmssm{} approach, respectively.
We make use of these effective couplings
both at \lo{} and \nlo{}, which results in $\Gamma^{\text{\lo{}}}_{\epsilon/\overline\epsilon}$
and $\Gamma^{\text{\nlo{}}}_{\epsilon/\overline\epsilon}(t,\tilde t)$, respectively.
For the latter to avoid double-counting of contributions,
we subtract the vertex correction using the exact value of
$\lambda^{\text{eff}}_{Hhh}(t,\tilde t)$ in $\mathcal{A}^{\text{eff}}$ in \eqn{eq:vertexle}.
This can be understood from the fact that the Feynman-diagrammatic
calculation of the partial decay width $H\to hh$
adds the exact corrections due to the top quark and the stops.
Lastly, in option 3 the calculation is performed in a
consistently matched \thdm{} to the \mssm{} with effective Higgs self-couplings
and subleading terms in the matching beyond the \hmssm{} approach.
This results in $\Gamma^{\text{\lo{}}}_{\text{eff}}$
and $\Gamma^{\text{\nlo{}}}_{\text{eff}}(t,\tilde t)$.
Both for option 2 and option 3 we use $\lambda^{\epsilon}_{ijk}$,
$\lambda^{\overline\epsilon}_{ijk}$ and
$\lambda^{\text{eff}}_{ijk}(1,1)$, respectively, in all occurrences of Higgs self-couplings
in \eqn{eq:Aext} and \eqn{eq:Aexteff}.
This choice is taking finite higher-order effects into account and does
not harm the cancellation of ultraviolet divergences.

\begin{table}
\begin{center}
\begin{tabular}{|c|cc|cc|}
  \hline
  Option & & $\mathcal{A}^{\text{\lo{}}}$ &  & $\mathcal{A}^{\text{eff}}$\\[0.1cm]
  \hline
  1 & $\Gamma^{\text{\lo{}}}$                     & $\lambda_{Hhh}$ & $\Gamma^{\text{\nlo{}}}(t,\tilde t)$ & 0 \\[0.1cm]
  2 & $\Gamma_{\overline\epsilon}^{\text{\lo{}}}$ & $\lambda^{\overline\epsilon}_{Hhh}$ & $\Gamma_{\overline\epsilon}^{\text{\nlo{}}}(t,\tilde t)$ & $-\Delta \lambda_{Hhh}(t,\tilde t)$\\[0.1cm]
  3 & $\Gamma_{\text{eff}}^{\text{\lo{}}}$        & $\lambda^{\text{eff}}_{Hhh}(1,1)$ & $\Gamma_{\text{eff}}^{\text{\nlo{}}}(t,\tilde t)$ & $-\Delta \lambda_{Hhh}(t,\tilde t)$\\[0.1cm]
  \hline
\end{tabular}
\end{center}
\caption{Different width calculations employed in our numerical analysis.\label{tab:widthoptions}}
\end{table}

\subsection{Comparison with earlier work}
Our method of calculating the decay width for $H\to hh$ deviates from
Ref.~\cite{Brignole:1992zv}, since the latter performed a $\overline{\rm DR}$
renormalization of all parameters involved, i.e.~did not include
an explicit decoupling of the heavy top-quark and stop states involving a
proper matching to the full \mssm{}. In this way the work of \citere{Brignole:1992zv}
could not isolate the pure momentum-dependent contributions
beyond the effective parameters rigorously. \s

The work of \citere{Williams:2007dc} presented the full \nlo{}
results within the complex \mssm{} with conventional \susy{}-electroweak
renormalization, i.e.~without the introduction of effective Higgs
self-couplings and mixing angles. This work finds large radiative
corrections that, however, should be explainable as the missing
contributions to the effective parameters at \lo{} to a large extent.
We will demonstrate this effect in our subsequent numerical analysis. Our
approach can be extended to the full calculation within the \mssm{}. This,
however, is beyond the scope of our work. \s

The recent work of \citere{Chalons:2017wnz} renormalized the Higgs
self-coupling in the $\overline{\rm MS}$-scheme so that also this work
did not perform an explicit matching of the low-energy \thdm{} to the full
\mssm{}. On the other hand the authors used the \hmssm{} approach to
approximate the dominant radiative corrections within the \mssm{} Higgs
sector. The residual effects beyond the use of their ``effective''
parameters, however, range at the same level as the consistent
momentum-dependent contributions beyond the effective couplings and
mixing angles as obtained in this work.

\section{Numerical Results \label{sec:numerical}}

We perform our comparisons in one benchmark scenario, in which we allow
for different values of the Higgsino mass parameter~$\mu$.
We will use the results of the work performed in \citere{Brucherseifer:2013qva}
for the effective Higgs self-couplings and masses at the one-loop level.
The most important parameter settings are as follows,
\bea
M_S=1500\,\GeV,\qquad X_t= \left\lbrace\begin{matrix}2950\,\GeV& \text{for }\tan\beta\leq 4\\
(2950 - \tfrac{400}{3}(\tan\beta - 4))\,\GeV &\text{for }\tan\beta>4\end{matrix}\right. \,,
\eea
where $M_S$ is the soft-\susy{} breaking mass used for all left- and right-handed squarks.
We set $\alpha_s(M_Z)=0.118$, $M_Z=91.15449$\,GeV, $m_t=173.2$\,GeV and
$G_F=1.166378\cdot~10^{-5}$\,GeV$^{-2}$. The latter fixes the vacuum expectation value~$v$.
The top-quark mass is understood to be on-shell and internally transformed into a $\overline{\text{DR}}$
top-quark mass at \nlo{} evaluated at $Q=M_S$. This transformation as well as the running
and matching of $\alpha_s$, see \citere{Brucherseifer:2013qva} for details,
induces a dependence on the gluino mass and the other squark masses,
which are fixed by the choices $M_3=2500$\,GeV, $m_b=4.84$\,GeV and $A_b=A_t$.
The quark masses of the first two generations are set to zero.
In the calculation of the Higgs masses and the effective Higgs self-couplings $\lambda_{ijk}^{\text{eff}}(1,1)$
at the one-loop level we also incorporate subleading $D$-terms in the squark masses
to match the results of \citere{Brucherseifer:2013qva}. The determination of the $D$-terms
also need the weak mixing angle, fixed through $M_Z$ above and $M_W=80.36951$\,GeV.
We checked that the inclusion of $D$-terms in the squark masses has almost no impact on our findings
due to the relatively heavy \susy{} scale, i.e.~$M_Z\ll M_S$. \s

The choice of $X_t$ allows to keep the light \cp{}-even Higgs mass close to $125$\,GeV at least
for values of $\tan\beta>4$. A Higgs mass of $125$\,GeV can also be reached for lower
values of $\tan\beta$, but only in combination with larger values of $M_S$. Larger values
motivate an RGE-running of the effective couplings. Since such a discussion is beyond
our scope, we stick to $M_S=1500$\,GeV and emphasize that our findings can be considered
quite general, despite the fact that we ``undershoot'' the experimental value of $125$\,GeV
significantly at low values of $\tan\beta$.
For a suitable and conclusive comparison we use the light \cp{}-even Higgs mass~$M_h$ obtained
in the \epa{} at the one-loop level as input to the \hmssm{} approach, see \eqn{eq:hMSSMapproach}.
This implies that our values of $\epsilon$ and $\overline\epsilon$ are based on a value
of $M_h$ that differs from $125$\,GeV for low values of $M_A$ or $\tan\beta$.
In the following we choose three different values of $\mu$.
Our first choice is $\mu=0$\,GeV, since only a Higgsino mass,
which is small compared to the stop masses,
allows the \hmssm{} approximation to be valid. We are aware that
$\mu=0$\,GeV results in very light Higgsinos,
which are not compatible with chargino mass bounds~\cite{PhysRevD.98.030001}.
Thus, the scenario with $\mu=0$\,GeV is considered pedagogical
to set up the consistency of our analysis with the \hmssm{} approach.
The second and third choice are $\mu=400$\,GeV and $\mu=1000$\,GeV. For the
latter choice, due to $\mu$ and $M_S$ being rather close, we will see
remaining differences between the \hmssm{} approximation and the exact one-loop results
in the Higgs masses, self-couplings and the partial decay width~$H\to hh$.
They arise, since we intrinsically violate an assumption of the
\hmssm{}, being $\mu \ll M_S$.

\subsection{Light Higgs mass and improved Higgs self-couplings}

\begin{figure}
  \begin{center}
    \begin{tabular}{ccc}
      \includegraphics[width=0.3\textwidth]{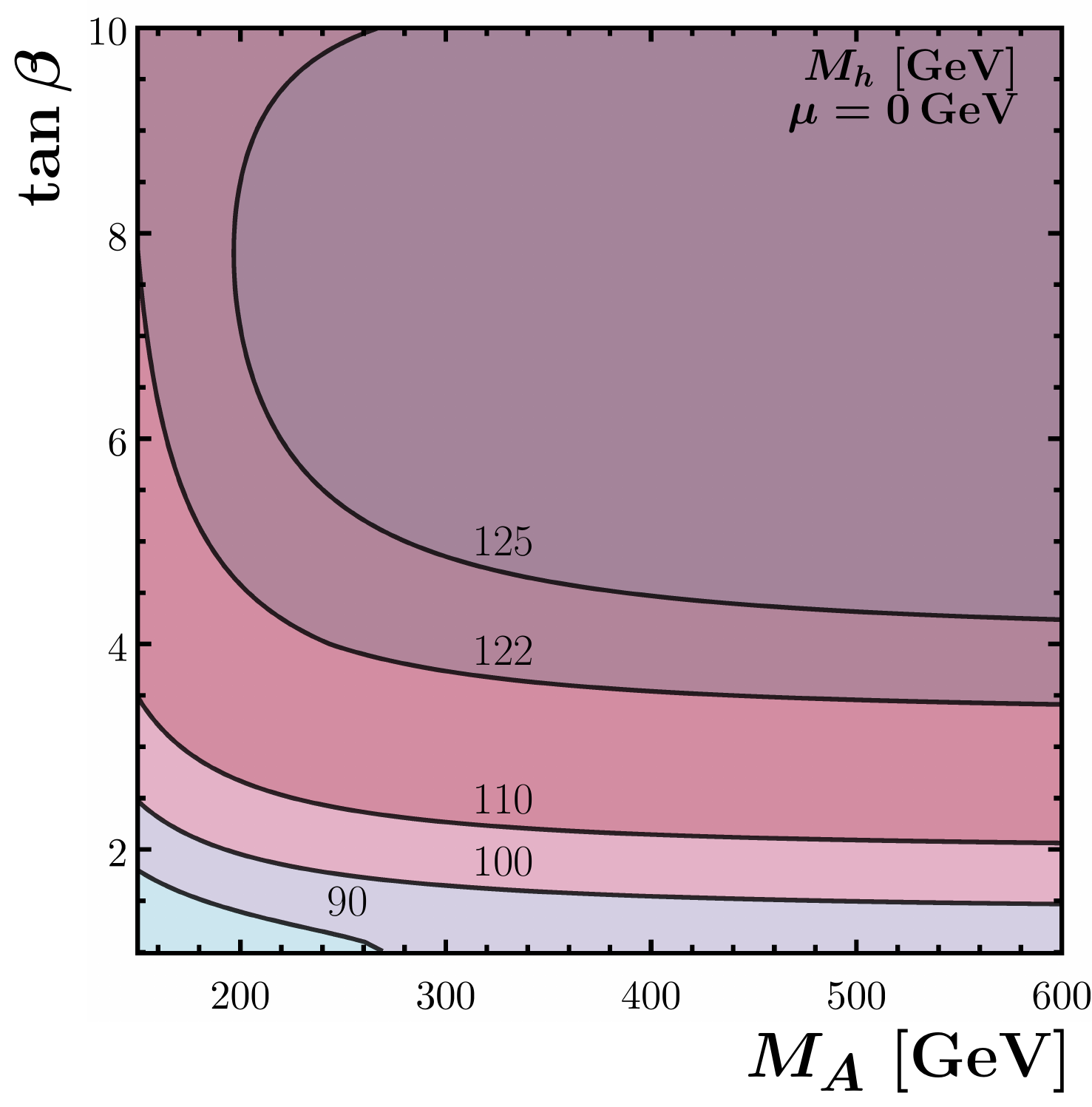} &
      \includegraphics[width=0.3\textwidth]{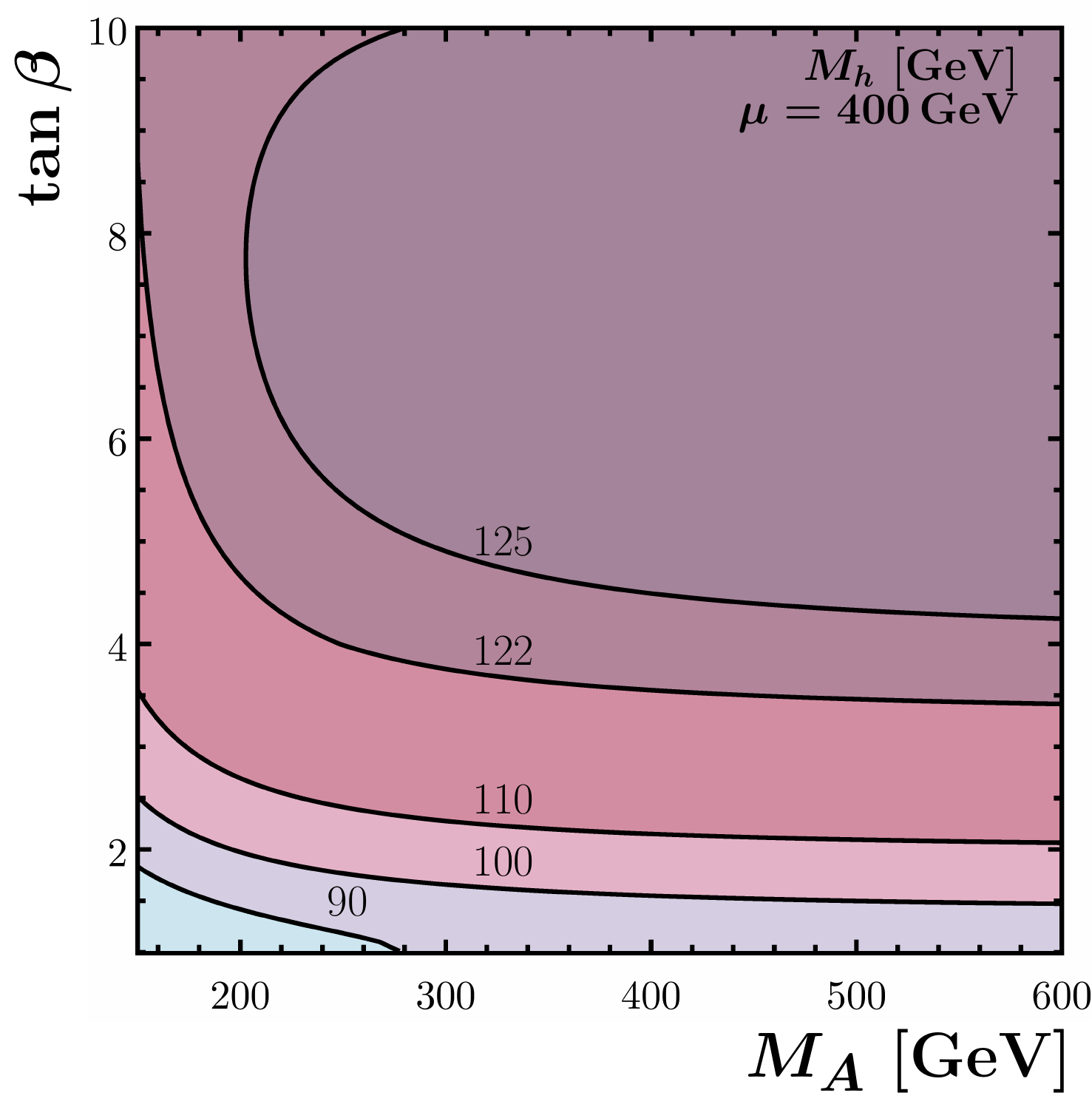} &
      \includegraphics[width=0.3\textwidth]{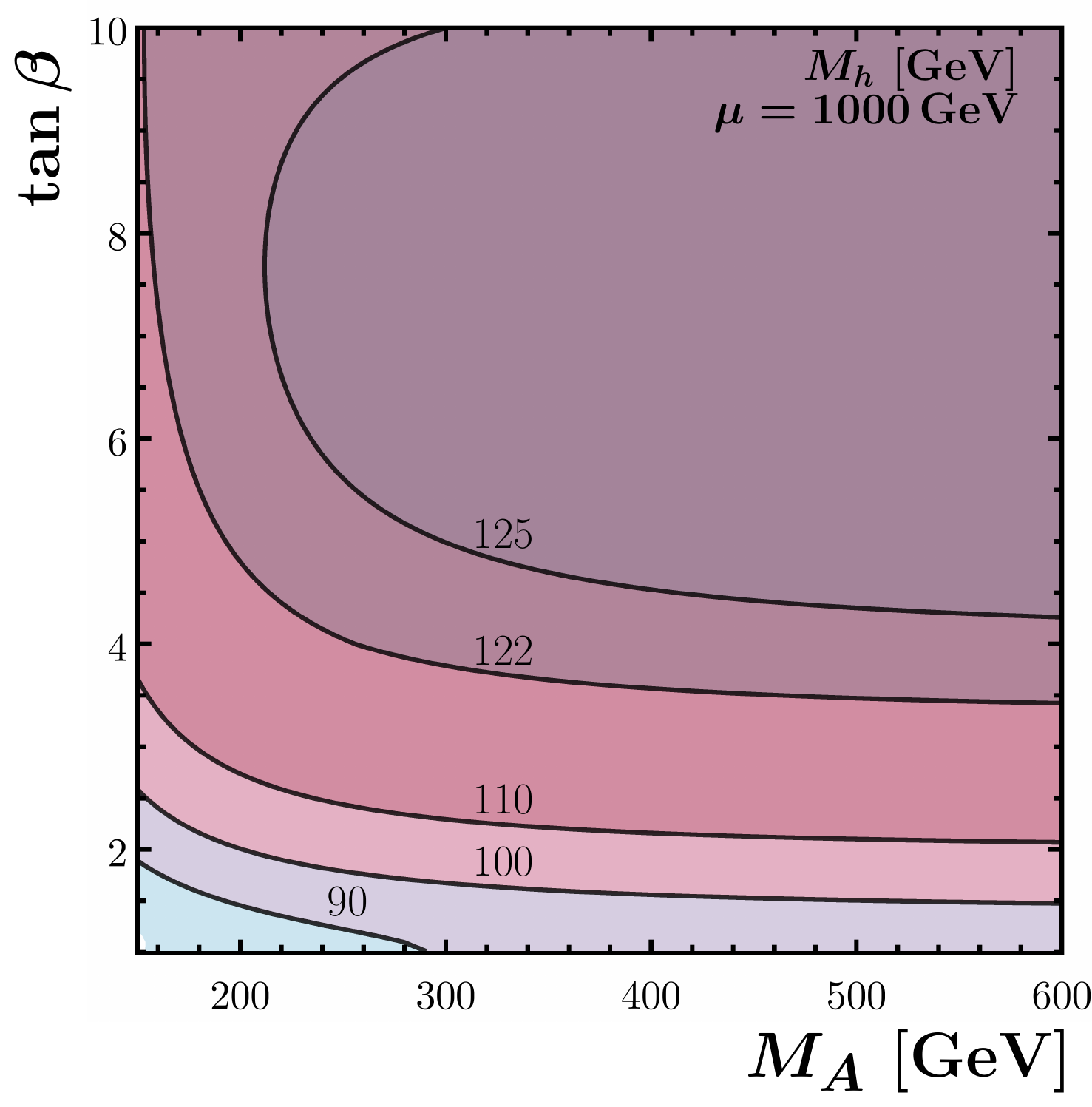}
      \\[-0.5cm]
      (a) & (b) & (c)
    \end{tabular}
    \parbox{\textwidth}{
      \caption[]{\label{fig:mLH} Light Higgs mass $M_h$ in GeV as a function of $M_A$ in GeV and $\tan\beta$ for (a) $\mu=0$\,GeV, (b) $\mu=400$\,GeV and (c) $\mu=1000$\,GeV.}}
  \end{center}
\end{figure}

\begin{figure}
  \begin{center}
    \begin{tabular}{cc}
      \includegraphics[width=0.4\textwidth]{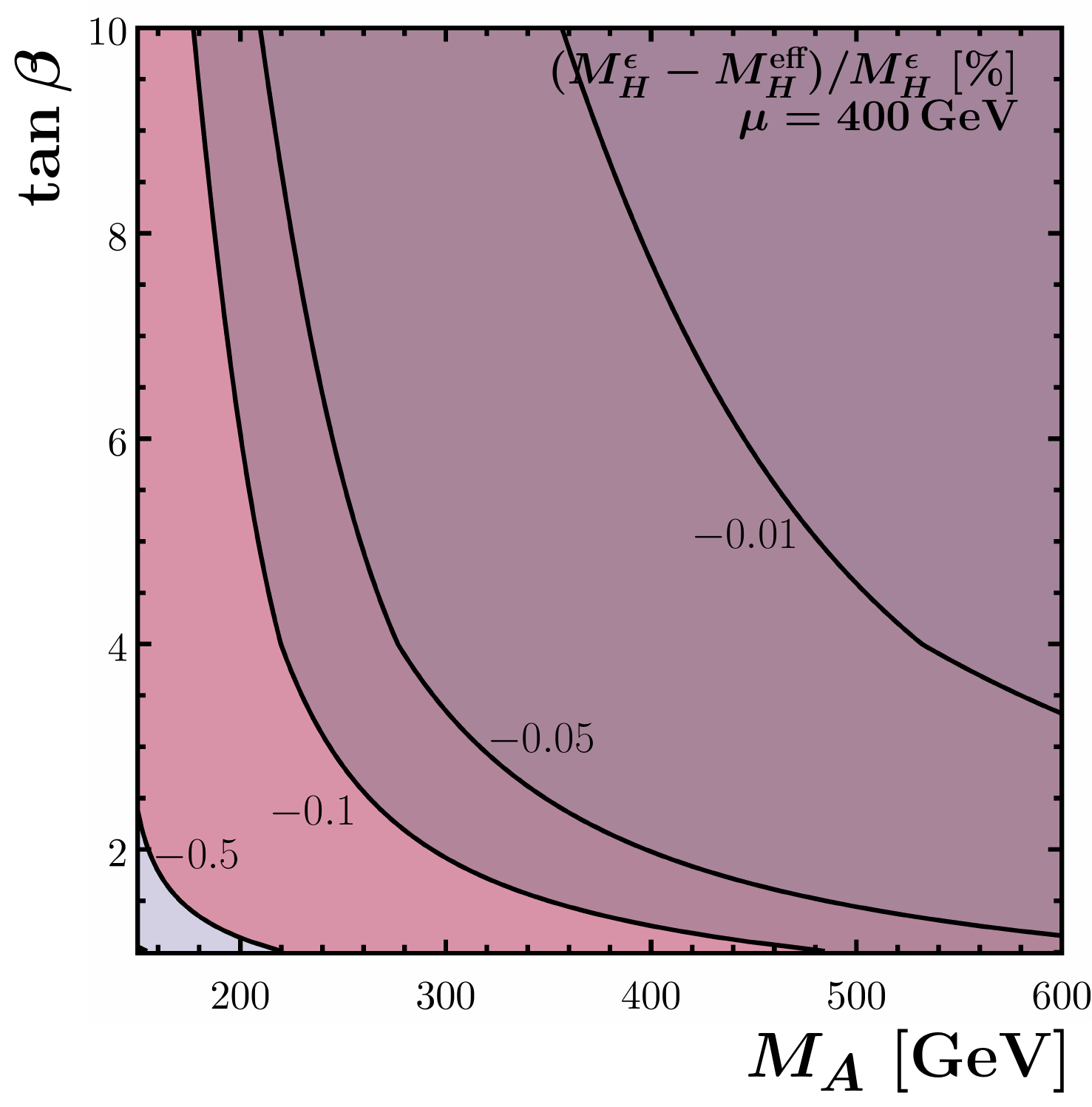} &
      \includegraphics[width=0.4\textwidth]{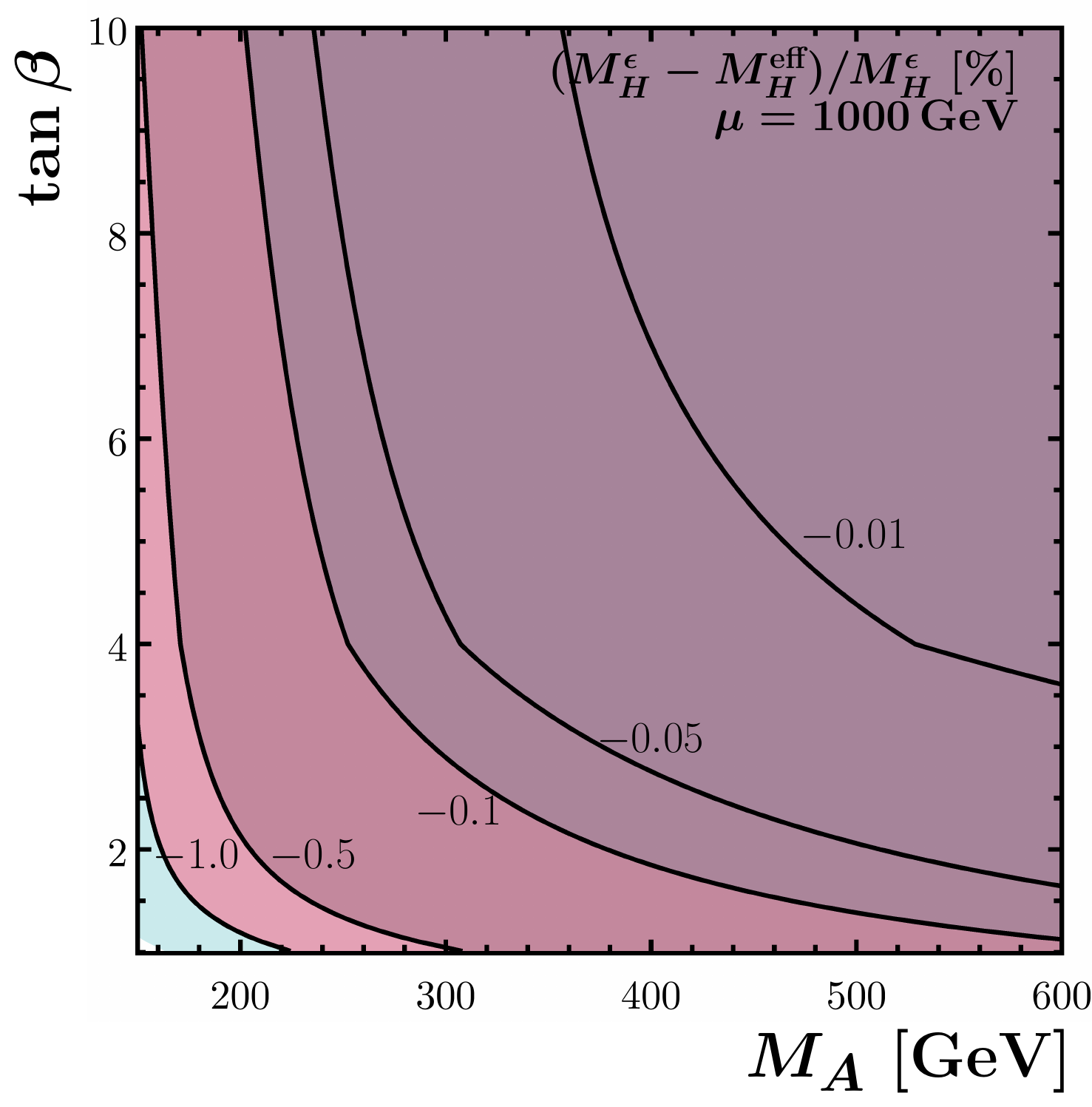} 
      \\[-0.5cm]
      (a) & (b) \\
      \includegraphics[width=0.4\textwidth]{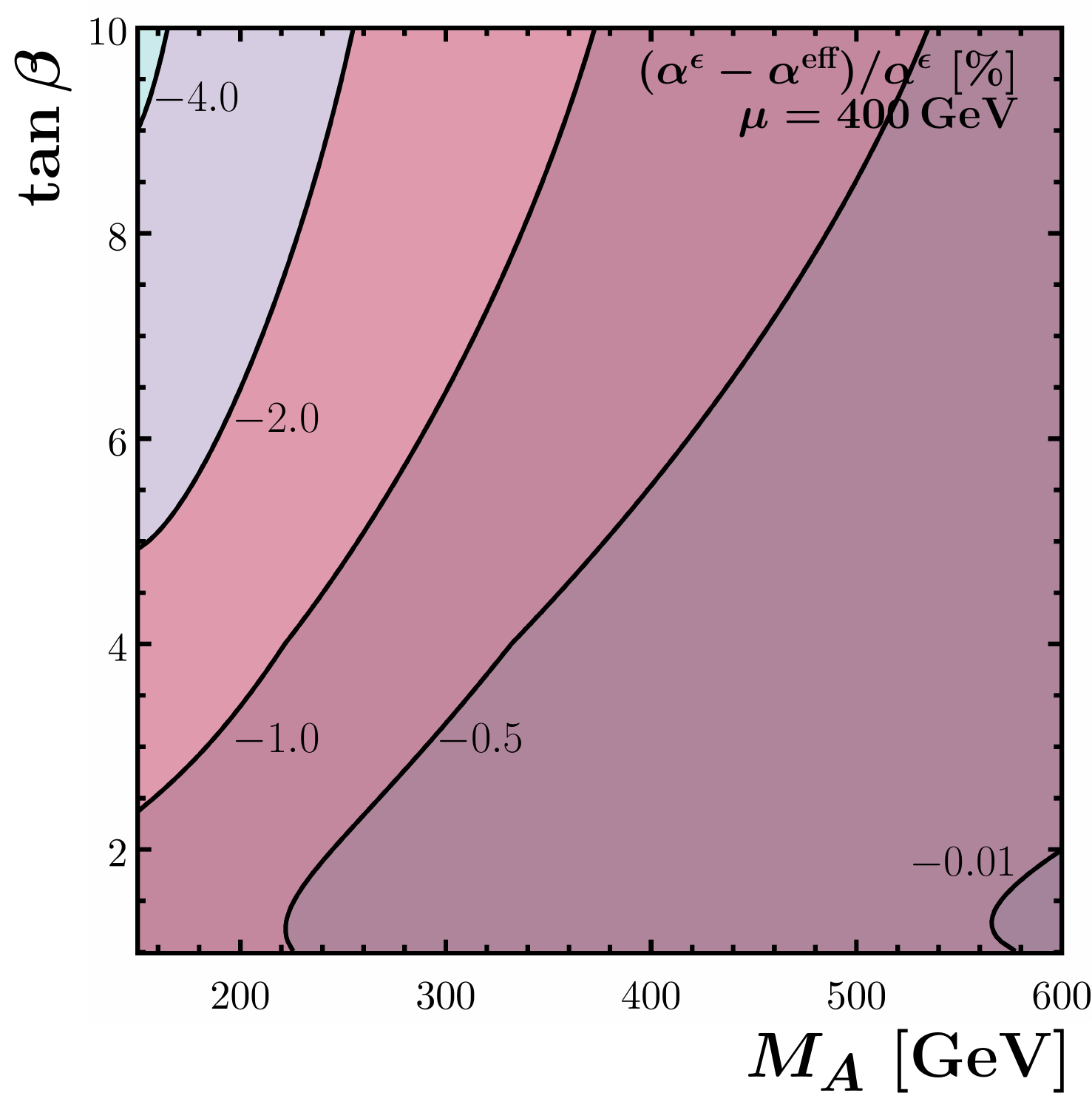} &
      \includegraphics[width=0.4\textwidth]{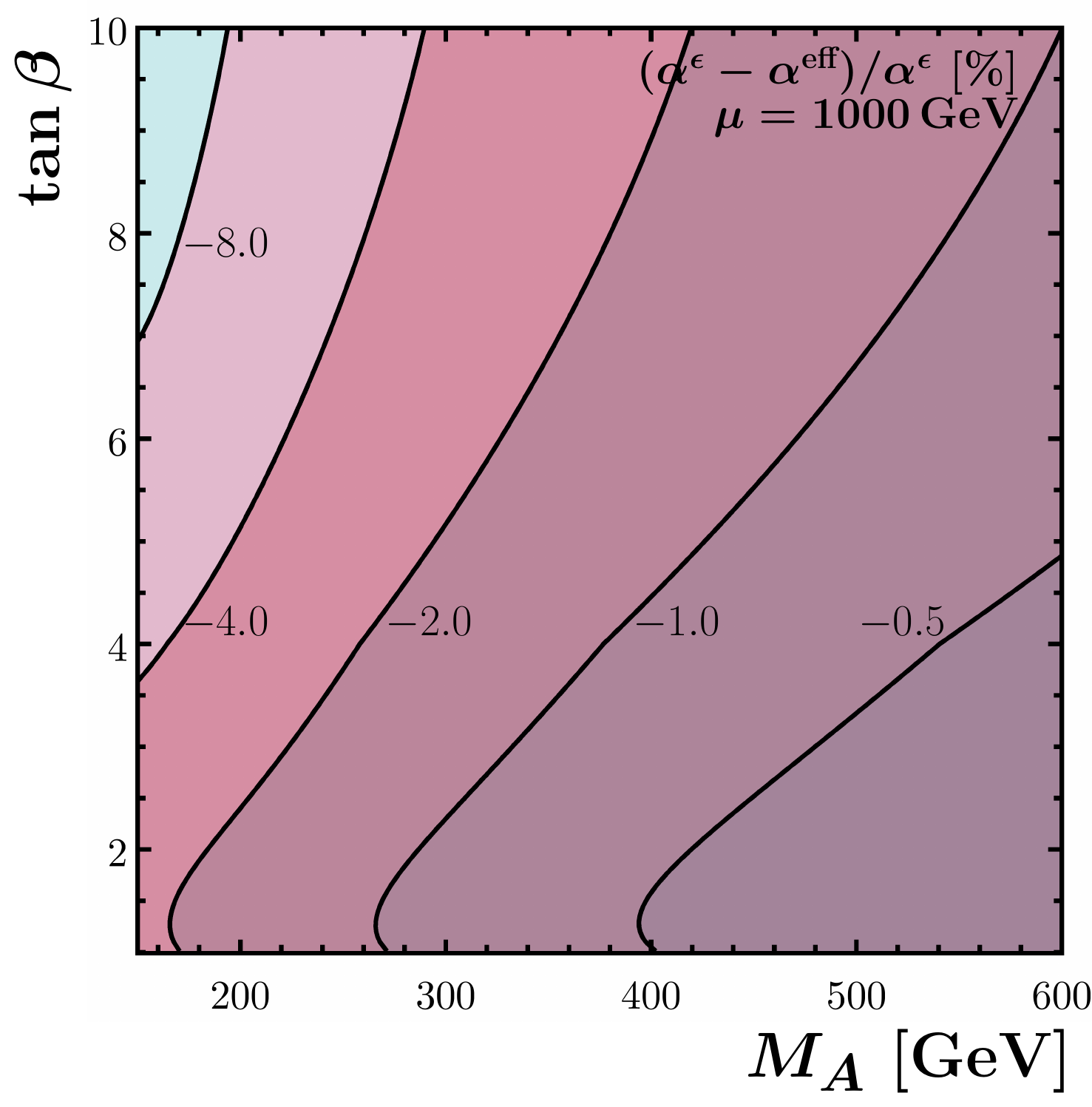} 
      \\[-0.5cm]
      (c) & (d) 
    \end{tabular}
    \parbox{\textwidth}{
      \caption[]{\label{fig:mHHalpha} Difference in the predictions of (a,b) the heavy Higgs-boson mass $M_H$ in \%
      and (c,d) the Higgs-boson mixing angle $\alpha$ in \%      
      between the predictions of the hMSSM obtained
      through \eqn{eq:HHmassepsilonapprox} and \eqn{eq:alphaepsilonapprox} and denoted $M_H^\epsilon$ and $\alpha^\epsilon$ with the exact values denoted $M_H^{\text{eff}}$ and $\alpha^{\text{eff}}$
      at \nlo{}, respectively,
      as a function of $M_A$ in GeV and $\tan\beta$ for (a,c) $\mu=400$\,GeV, (b,d) $\mu=1000$\,GeV. For $\mu=0$\,GeV such differences vanish.}}
  \end{center}
\end{figure}

We display the value of the light Higgs mass $M_h$ in \fig{fig:mLH} for three values of
$\mu=0$\,GeV, $400$\,GeV and $1000$\,GeV. The Higgs mass is calculated at \nlo{} taking
into account the corrections described in \sct{sec:higgsmasscorr}.
Since the value of $X_t$ is equal in all scenarios and
thus are the stop masses, the light Higgs mass is almost identical in the three scenarios.
We emphasize that for $\mu=0$\,GeV the exact values of the heavy Higgs mass $m_H$
and the Higgs mixing angle $\alpha$ can be obtained through 
\eqn{eq:HHmassepsilonapprox} and \eqn{eq:alphaepsilonapprox}.
For $\mu=400$\,GeV and $\mu=1000$\,GeV we depict the differences between
the predictions of the hMSSM approach through \eqn{eq:HHmassepsilonapprox} and \eqn{eq:alphaepsilonapprox}
and their exact determination at \nlo{} in \fig{fig:mHHalpha}.
For the heavy Higgs mass such differences are mostly below $1\%$ throughout the
parameter plane. For the Higgs mixing angle $\alpha$ instead differences rise
to a couple of percent, in particular for a very large value of $\mu=1000$\,GeV,
which is close to the soft-\susy{} breaking mass $M_S=1500$\,GeV in our example.
\citere{Bagnaschi:2015hka} showed
partially smaller discrepancies for the Higgs mixing angle $\alpha$ due to the smaller
ratio $\mu/M_S\ll 1$. We emphasize again that for $\mu=0$\,GeV we find perfect agreement
between the two approaches. Thus, we note that such discrepancies in $m_H$ and $\alpha$
are not the dominant source of differences observed in the partial decay width of $H\to hh$
in previous studies. On the other hand, as for the subsequently discussed Higgs self-couplings,
the differences in $\alpha$ for larger $\mu/M_S$ point towards the limitations of the hMSSM approach, see below. \s

\begin{figure}
  \begin{center}
    \begin{tabular}{ccc}
      \includegraphics[width=0.3\textwidth]{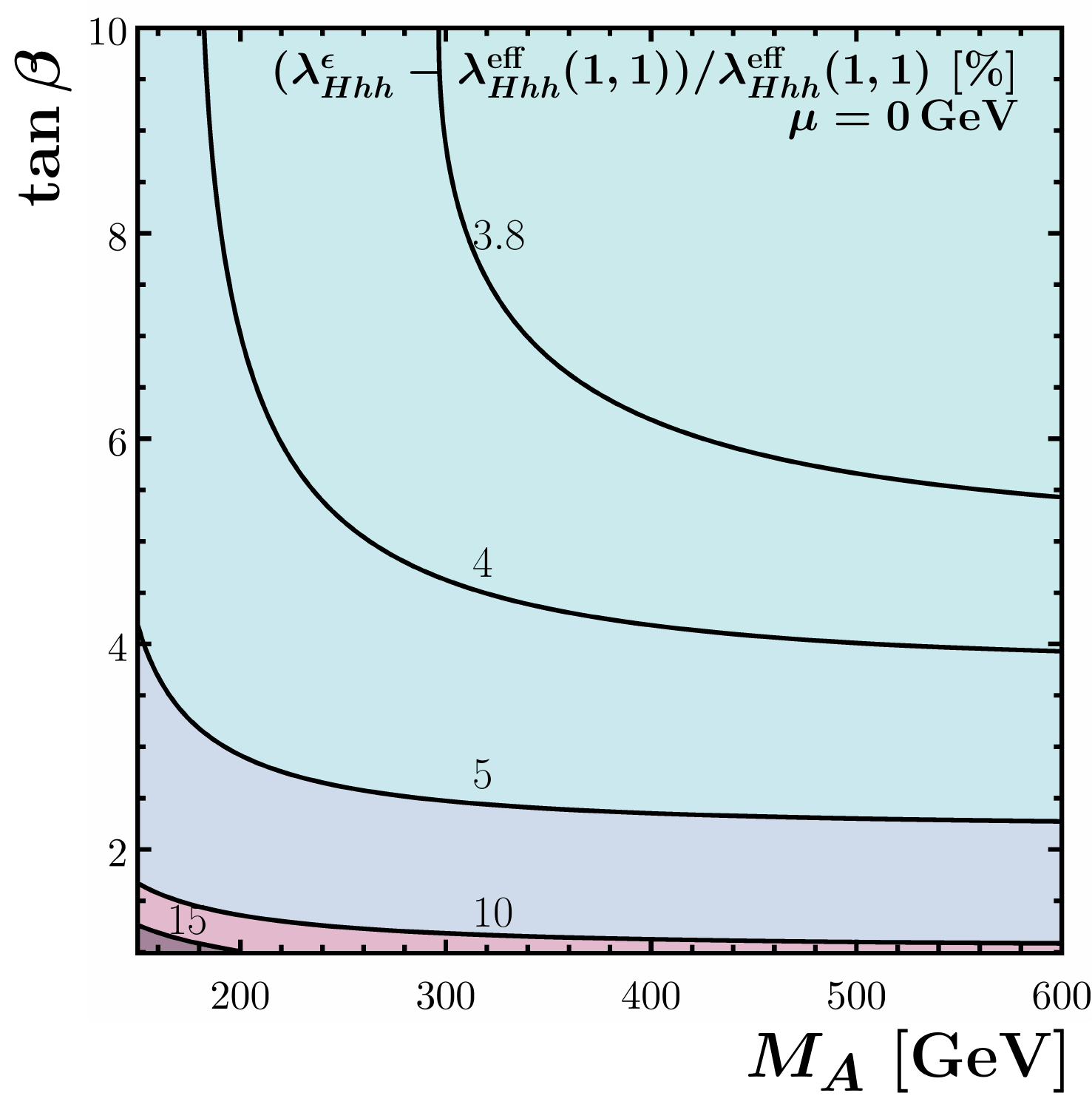} &
      \includegraphics[width=0.3\textwidth]{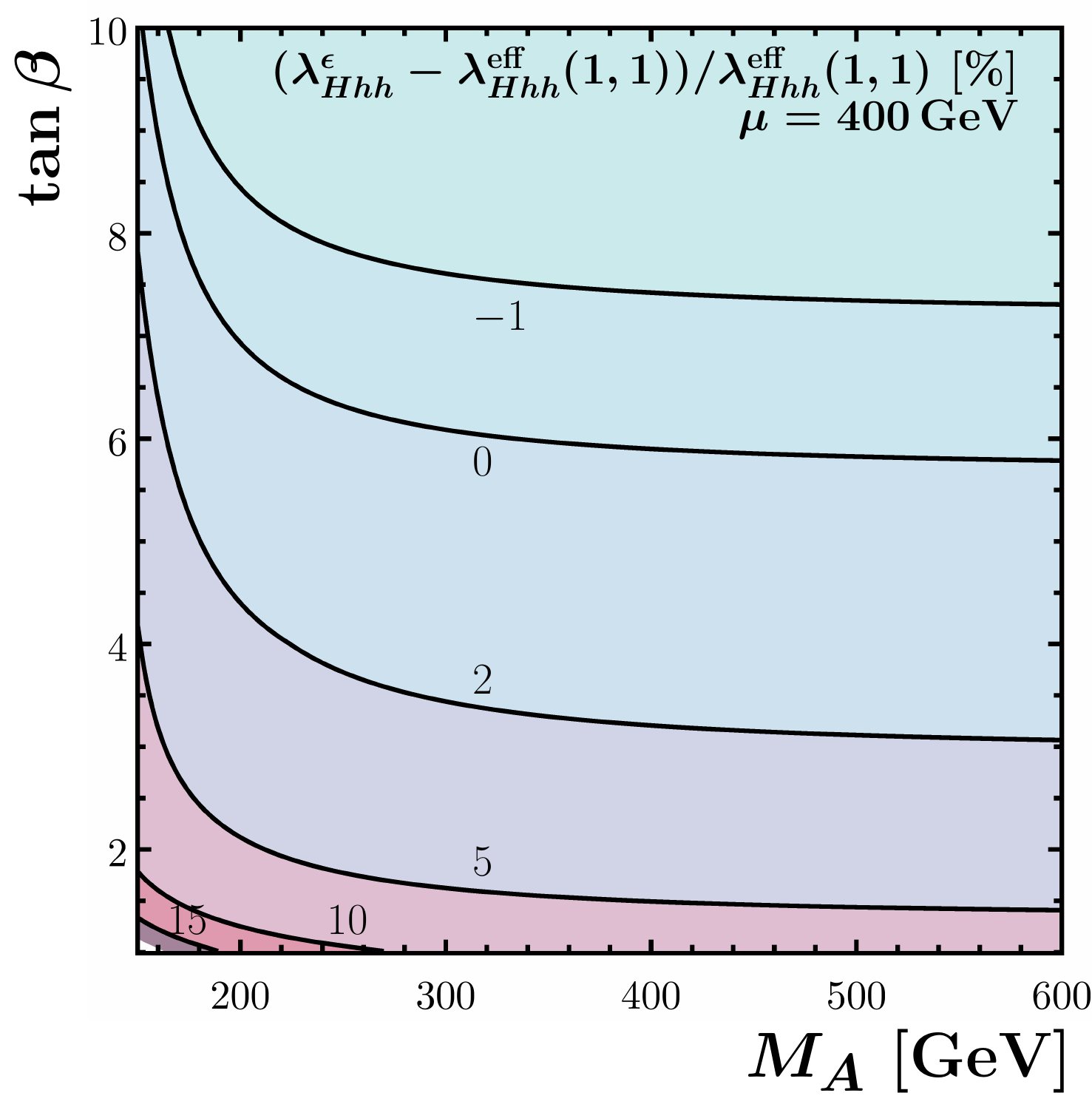} &
      \includegraphics[width=0.3\textwidth]{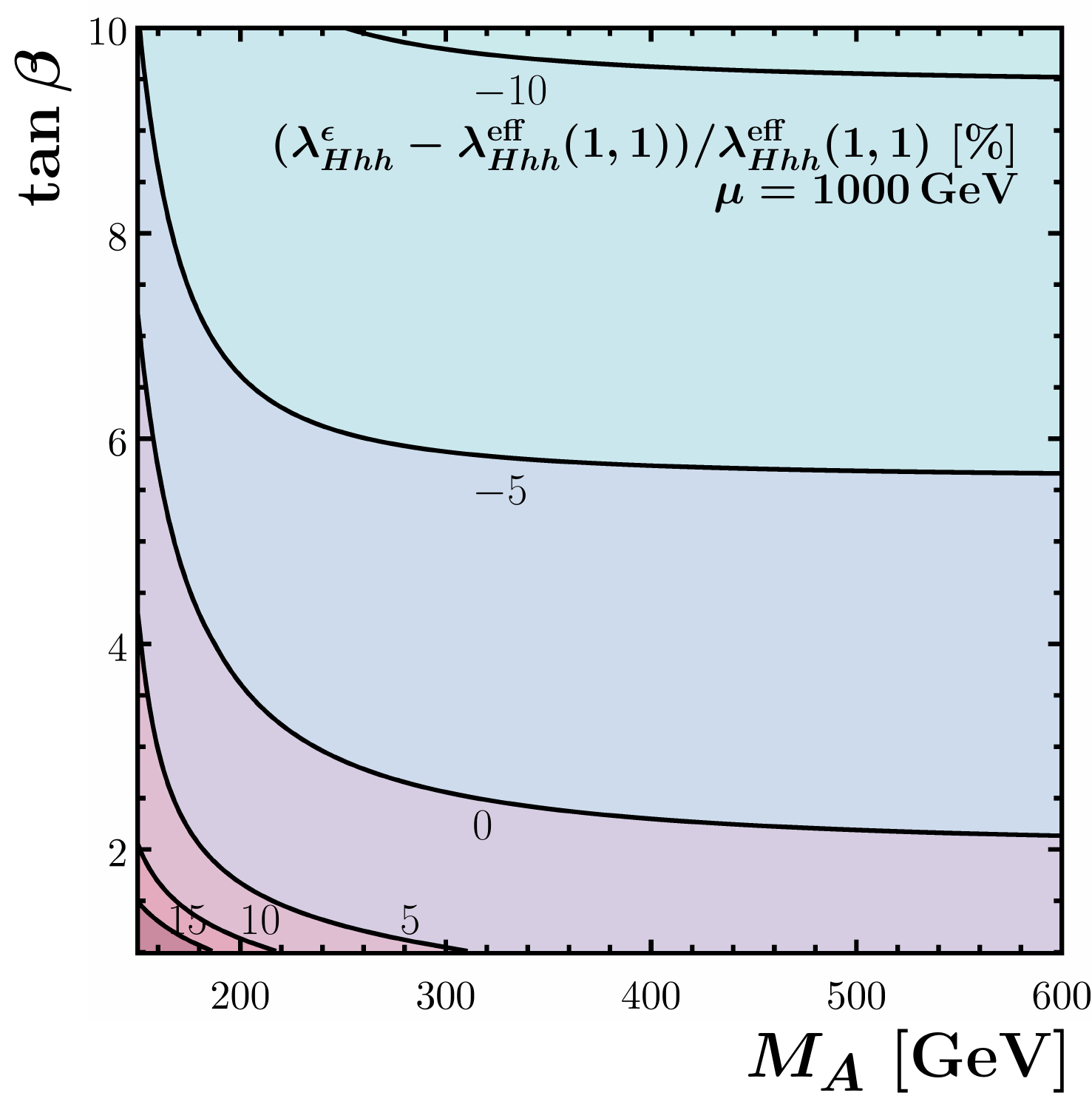}
      \\[-0.5cm]
      (a) & (b) & (c)\\
      \includegraphics[width=0.3\textwidth]{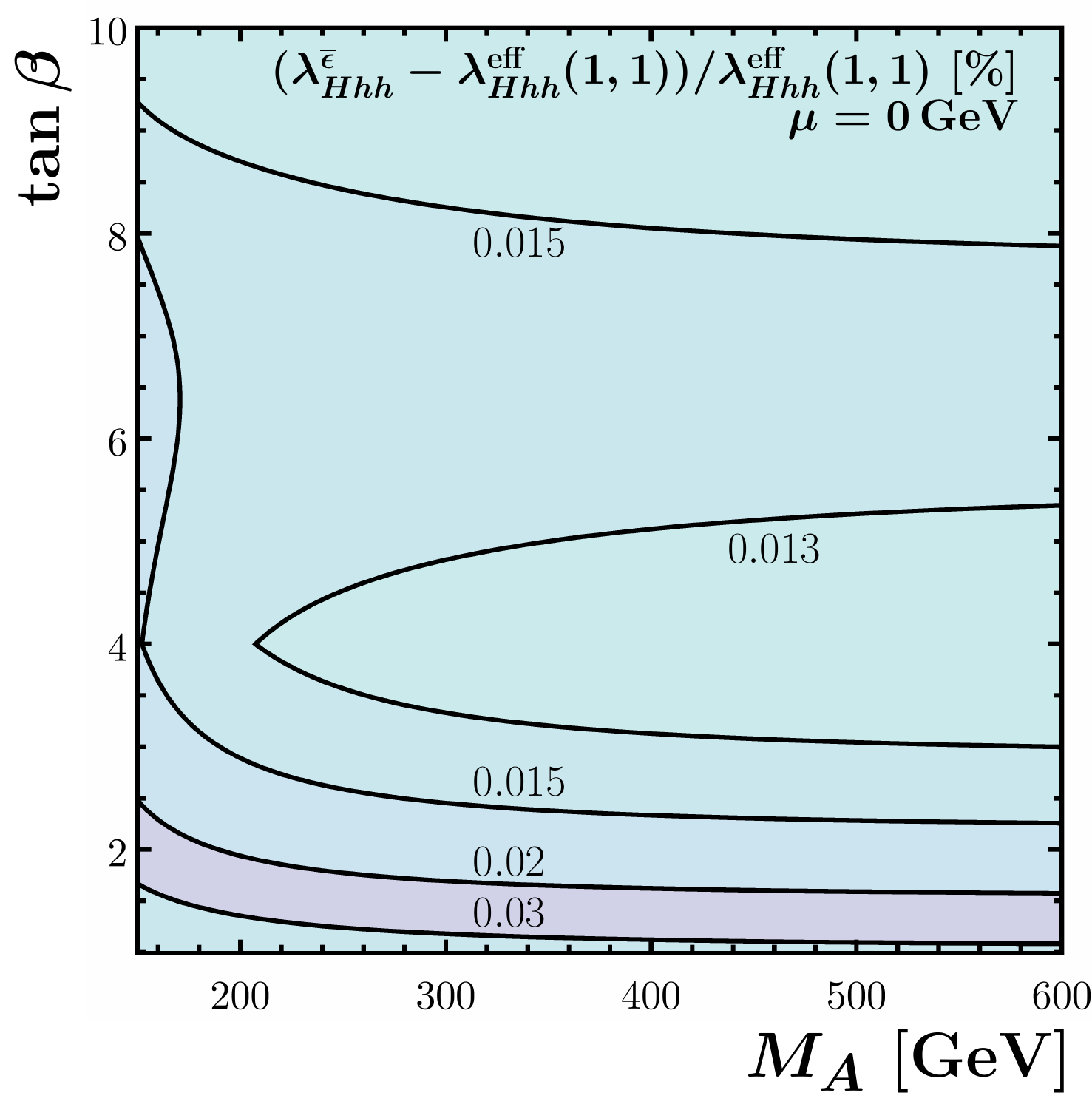} &
      \includegraphics[width=0.3\textwidth]{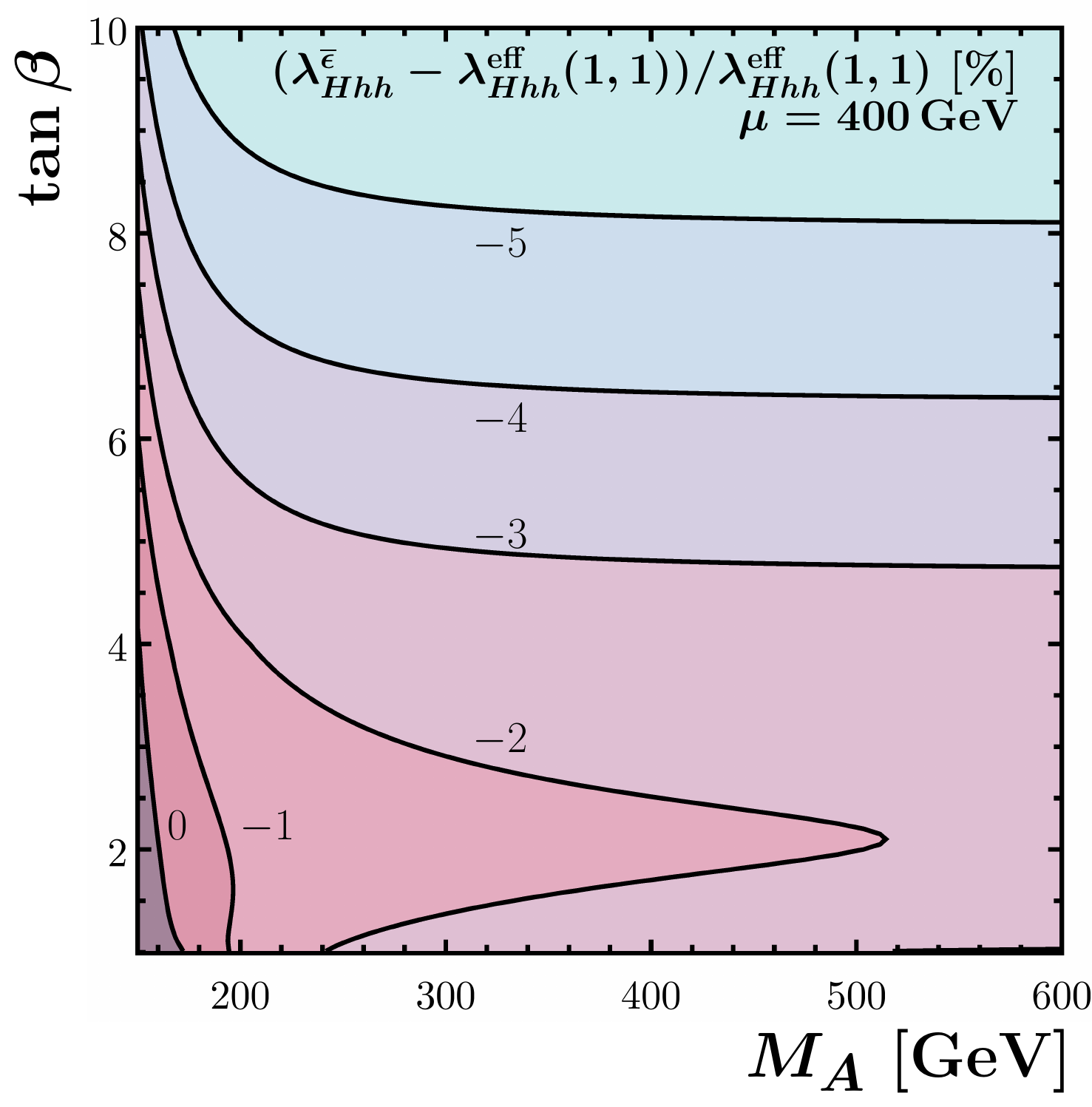} &
      \includegraphics[width=0.3\textwidth]{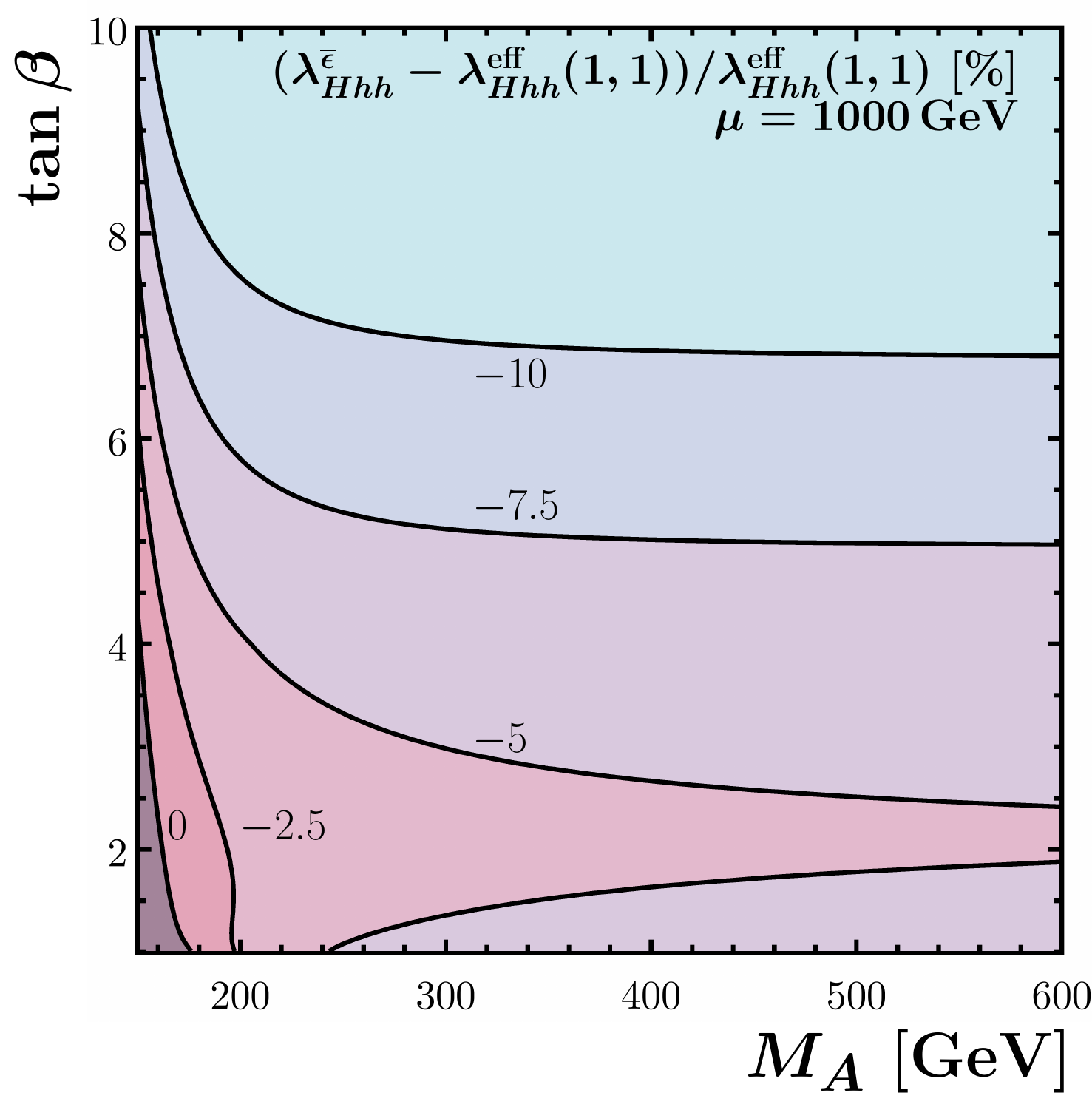}
      \\[-0.5cm]
      (d) & (e) & (f)
    \end{tabular}
    \parbox{\textwidth}{
      \caption[]{\label{fig:lambda} (a,b,c) $\lambda_{Hhh}^\epsilon$ and (d,e,f) $\lambda_{Hhh}^{\overline\epsilon}$ relative to the exact value $\lambda_{Hhh}^{\text{eff}}(1,1)$
      as a function of $M_A$ in GeV and $\tan\beta$ for (a,d) $\mu=0$\,GeV, (b,e) $\mu=400$\,GeV and (c,f) $\mu=1000$\,GeV.}}
  \end{center}
\end{figure}

We show the Higgs self-coupling $\lambda_{Hhh}$ in \fig{fig:lambda} again
for $\mu=0$\,GeV, $400$\,GeV and $1000$\,GeV. While the Higgs self-coupling obtained
in the standard \hmssm{} approach $\lambda^{\epsilon}_{Hhh}$ shows large deviations
from the exact value $\lambda^{\text{eff}}_{Hhh}(1,1)$, the improved version $\lambda^{\overline\epsilon}_{Hhh}$,
performs significantly better, compare \fig{fig:lambda}~(a) and (d).
The kinks at $\tan\beta=4$ in \fig{fig:lambda}~(d) 
are induced by our choice of $X_t$, which is constant for $\tan\beta \leq 4$ and rescaled for $\tan\beta>4$.
For non-vanishing $\mu$ the improved coupling shows remaining differences of a few percent,
see \fig{fig:lambda}~(e) and (f).
This is not surprising, since a non-vanishing $\mu$ violates an assumption
of the \hmssm{} approach: For $\mu\neq 0$\,GeV all entries in the correction
of the \cp{}-even Higgs mass matrix in \eqn{eq:corrmassmssm} receive corrections.
Still, we conclude that the improved couplings $\lambda^{\overline\epsilon}_{Hhh}$
compared to $\lambda^{\epsilon}_{Hhh}$ show a milder spread when compared
to the exact coupling $\lambda^{\text{\nlo{}}}_{Hhh}(1,1)$ throughout
the $M_A$-$\tan\beta$-plane for all values of $\mu$.

\subsection{Momentum-dependent and kinetic corrections to $H\to hh$}
\begin{figure}
  \begin{center}
    \begin{tabular}{ccc}
      \includegraphics[width=0.3\textwidth]{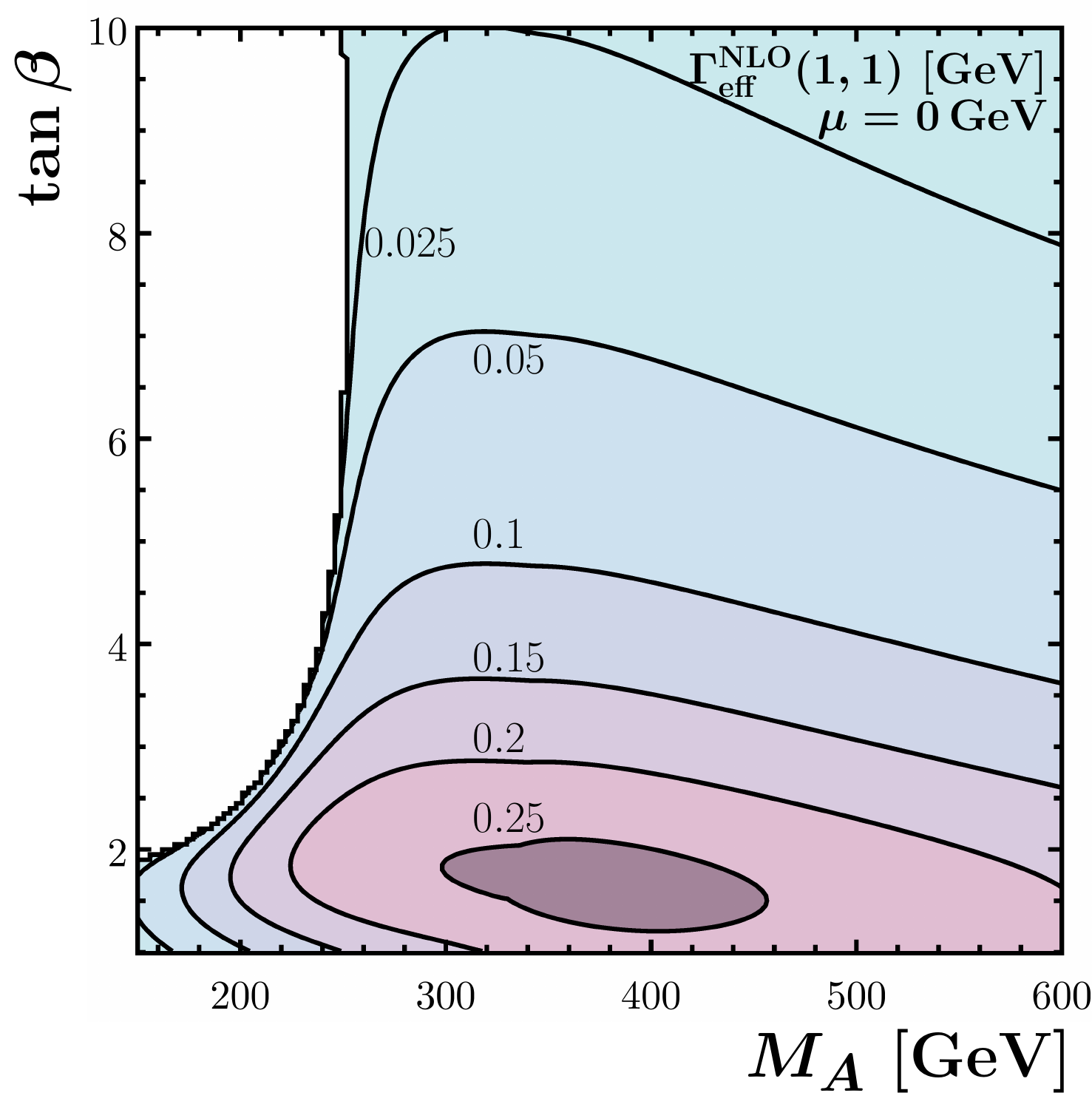} &
      \includegraphics[width=0.3\textwidth]{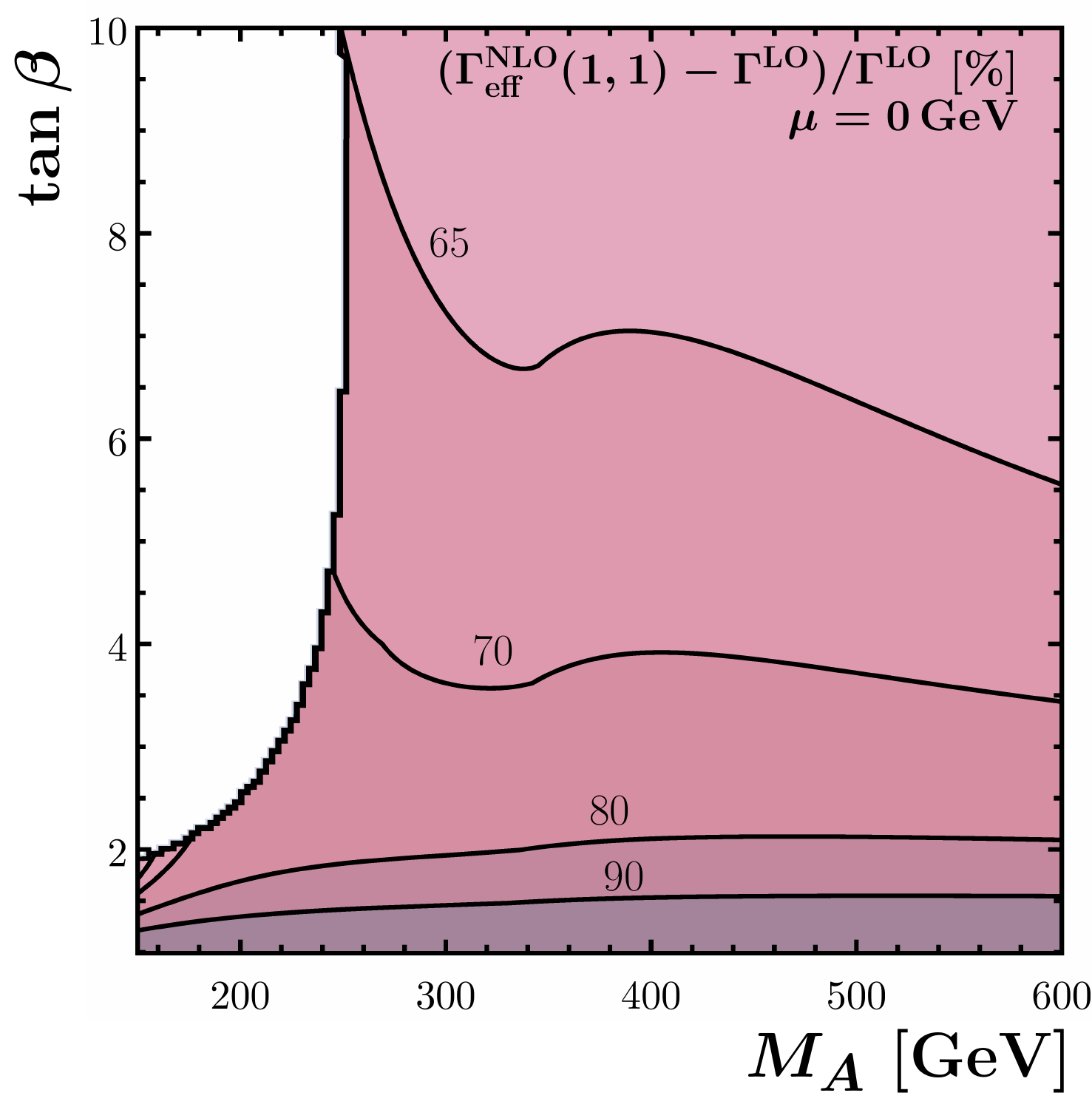} & 
      \includegraphics[width=0.3\textwidth]{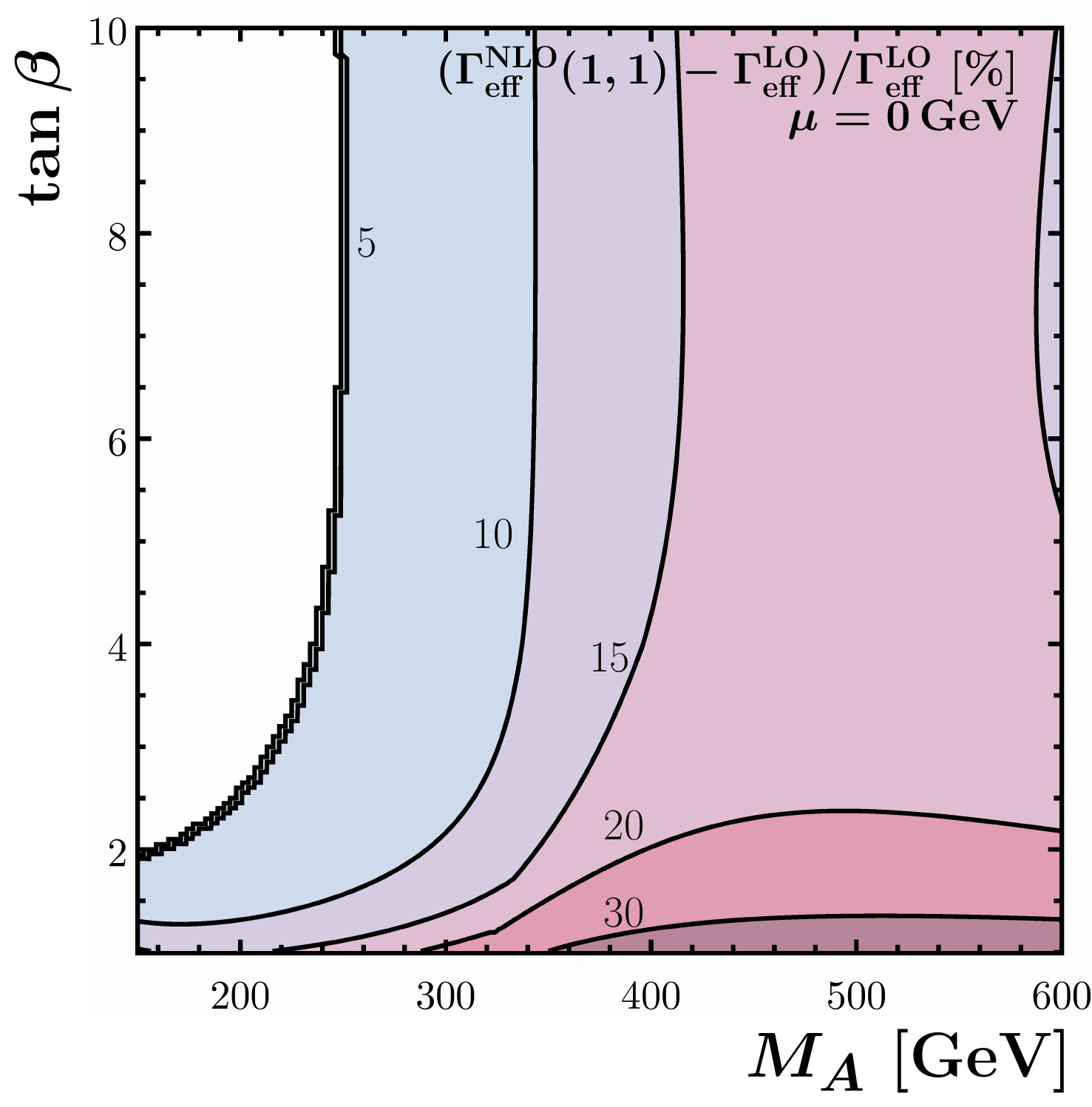}      
      \\[-0.5cm]
      (a) & (b) & (c)
    \end{tabular}
    \parbox{\textwidth}{
      \caption[]{\label{fig:Gammatotal} (a) Partial decay width $H\to hh$ in GeV at NLO as a function of $M_A$ and $\tan\beta$; (b) Relative correction with respect to the pure tree-level result $\Gamma^{\text{\lo{}}}$;
      (c) Relative correction with respect to the tree-level result $\Gamma^{\text{\lo{}}}_{\text{eff}}$ including $\lambda^{\text{eff}}_{Hhh}(1,1)$.
      }}
  \end{center}
\end{figure}
Before we discuss the quality of the \hmssm{} approximation for what concerns $H\to hh$
we focus on the relevance of momentum-dependent and kinetic corrections for the partial decay width $H\to hh$
in the benchmark scenario with $\mu=0$\,GeV. They arise from our Feynman-diagrammatic calculation with non-zero external
momenta and the kinetic mixing discussed in \sct{sec:efflag}, respectively.
The total partial decay width for $H\to hh$, $\Gamma^{\text{\nlo{}}}_{\text{eff}}(1,1)$,
including all top-quark and stop corrections is depicted in \fig{fig:Gammatotal}~(a). First it is apparent that for
low values of $M_A$ the heavy Higgs boson~$H$ is too light to decay into a pair of two \cp{}-even light Higgs bosons $hh$,
such that the on-shell decay is kinematically closed. Only for low values of $\tan\beta$, where $M_h$ is decreasing rapidly,
this decay mode is again of relevance. In all subsequent figures the region,
where the on-shell decay $H\to hh$ is kinematically closed, is shown in white.
We show the \nlo{} partial decay width $\Gamma^{\text{\nlo{}}}_{\text{eff}}$ relative
to the pure tree-level partial width $\Gamma^{\text{\lo{}}}$ including the tree-level Higgs self-coupling $\lambda_{Hhh}$
in \fig{fig:Gammatotal}~(b). We observe large corrections in accordance with the literature, see e.g. \citere{Williams:2007dc}.
However, we point out that using the effective coupling $\lambda^{\text{eff}}_{Hhh}(1,1)$ in the tree-level decay width $\Gamma^{\text{\lo{}}}_{\text{eff}}$
diminishes the one-loop corrections substantially, see \fig{fig:Gammatotal}~(c).
Accordingly, \fig{fig:Gammatotal}~(c) demonstrates the relevance of the remaining momentum-dependent and kinetic corrections,
which we incorporated through a Feynman-diagrammatic calculation and through the inclusion of the external kinetic mixing
described in \sct{sec:efflag}, respectively.
Exactly such momentum-dependent and kinetic contributions were missing in the comparison of $H\to hh$ decay widths
performed in \citere{Bagnaschi:2015hka}: Therein the partial decay width $\Gamma^{\text{\lo{}}}_{\epsilon}$
was compared against a full Feynman-diagrammatic one-loop calculation including an additional resummation of large logarithms.
The discrepancies found at the level of the partial decay width in \citere{Bagnaschi:2015hka} were 
between $15-25$\% in a large part of the $M_A$-$\tan\beta$-plane
and thus in the same ballpark as the missing momentum-dependent and kinetic contributions. 
Keep in mind that the partial decay width~$\Gamma^{\text{\lo{}}}_{\epsilon}$
and thus the comparison in \citere{Bagnaschi:2015hka} missed also the constant factor between $\epsilon$ and $\overline\epsilon$.
We leave a detailed
comparison employing the newest predictions for a future study.
We emphasize that the momentum-dependent and kinetic
corrections in our benchmark scenario are mostly induced by the top-quark contribution,
whereas the stops only yield a tiny fraction due to the relatively
large stop masses.
Given that in \fig{fig:Gammatotal} we are not yet applying the \hmssm{} approximation
similar results are obtained for larger values of $\mu$.
We therefore refrain from showing the corresponding results for $\mu=400$\,GeV or $\mu=1000$\,GeV.

\subsection{$H\to hh$ in the hMSSM}
\begin{figure}
  \begin{center}
    \begin{tabular}{ccc}
      \includegraphics[width=0.3\textwidth]{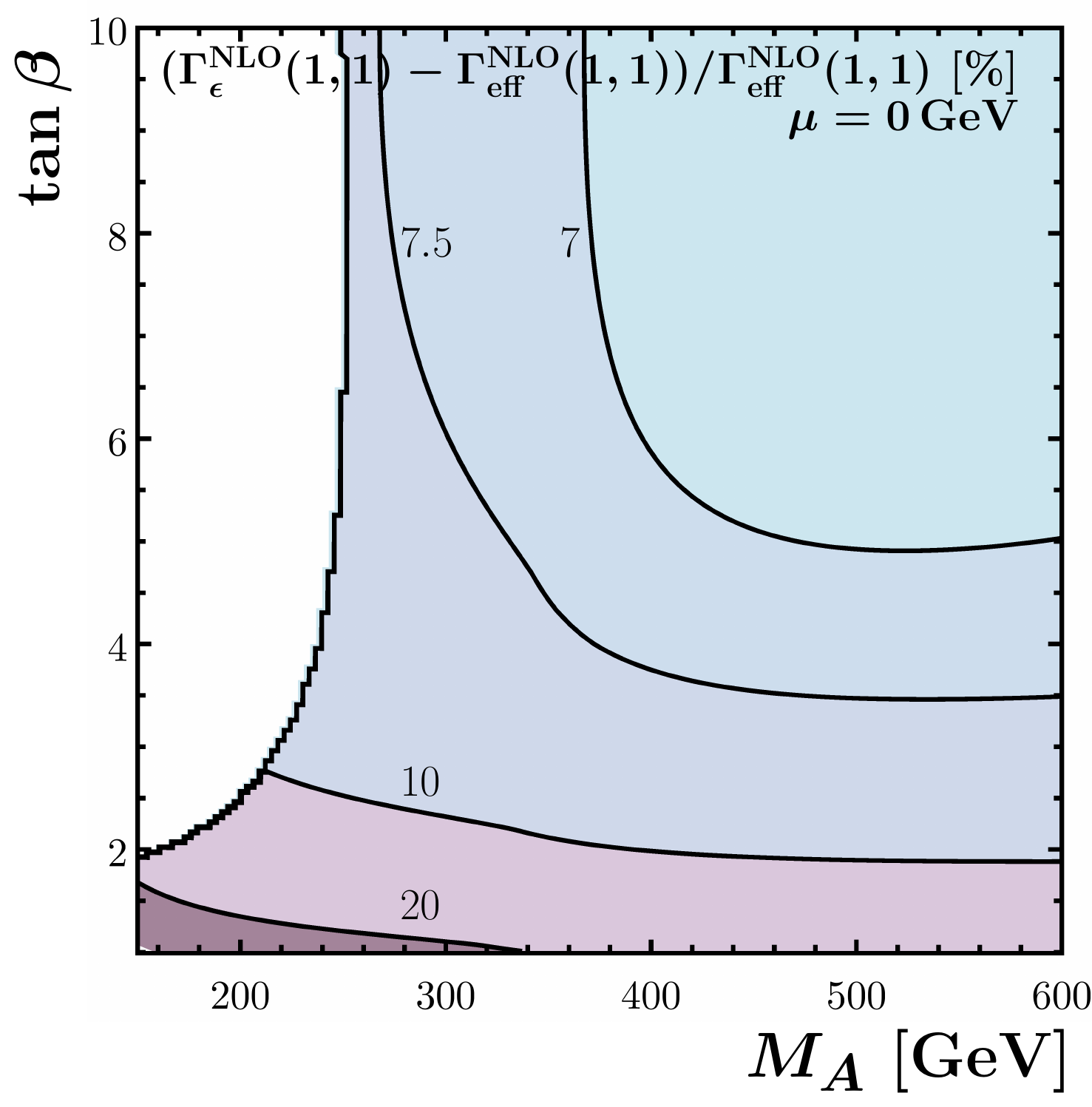} &
      \includegraphics[width=0.3\textwidth]{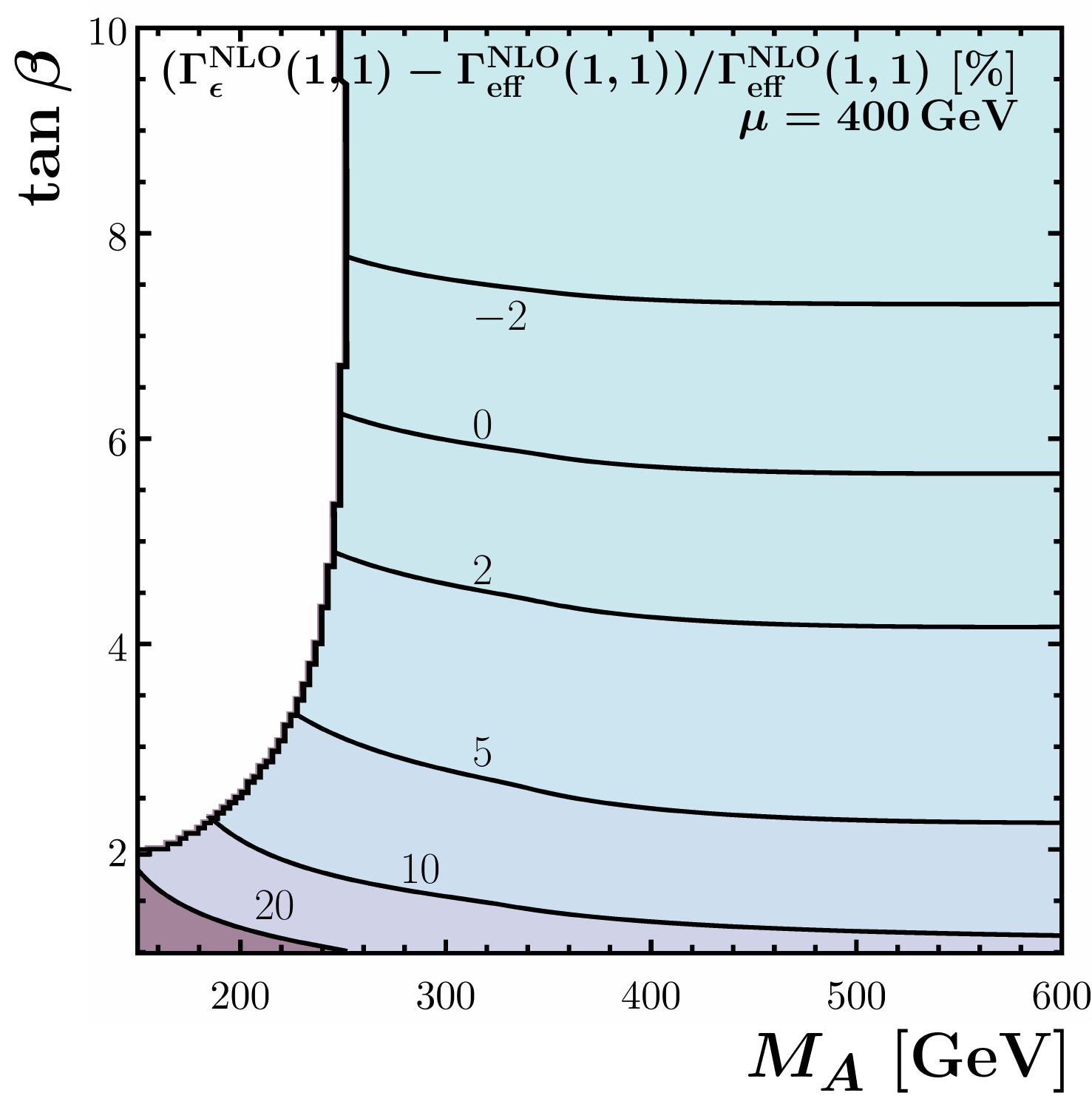} &
      \includegraphics[width=0.3\textwidth]{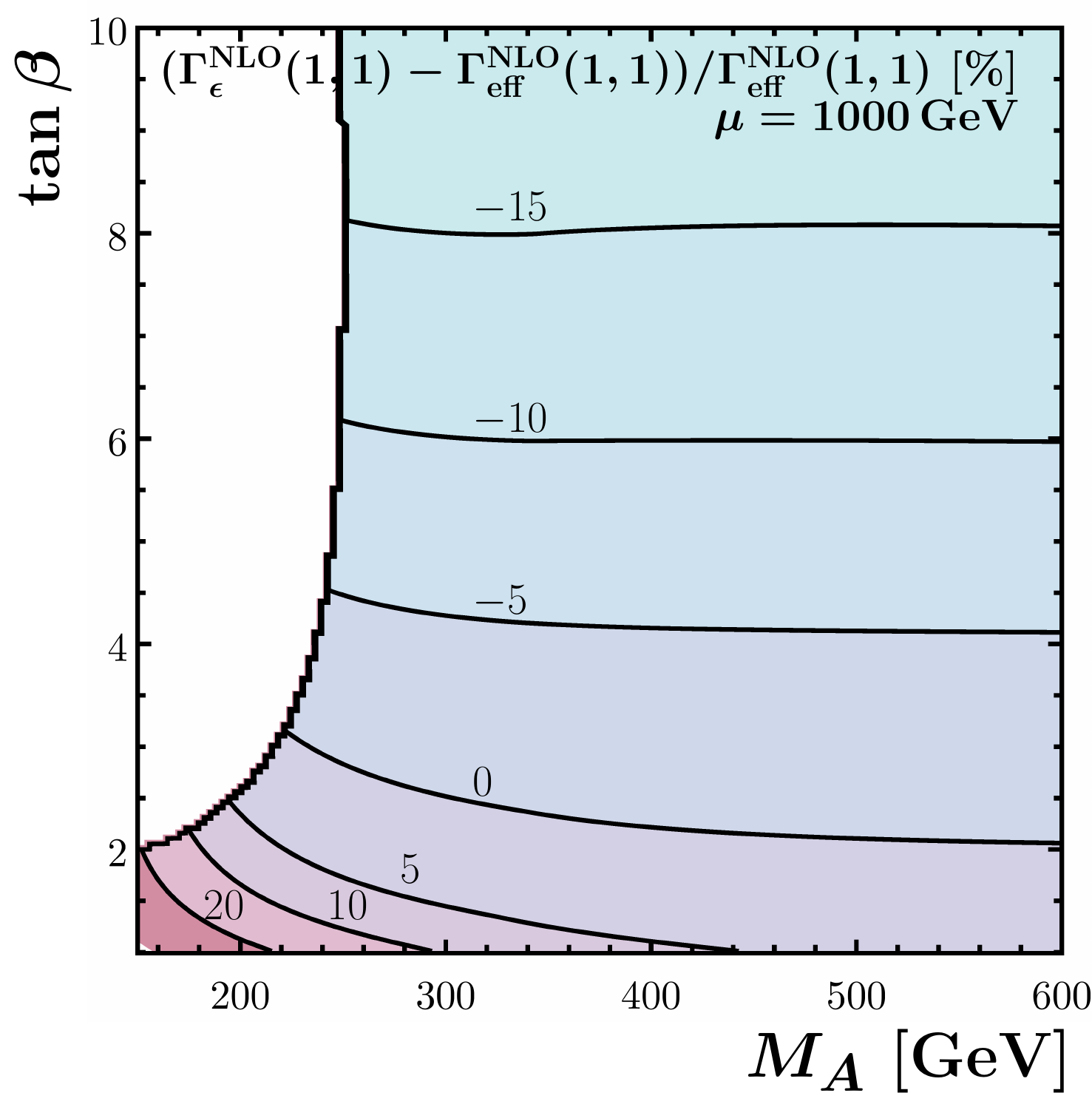}      
      \\[-0.5cm]
      (a) & (b) & (c)\\
      \includegraphics[width=0.3\textwidth]{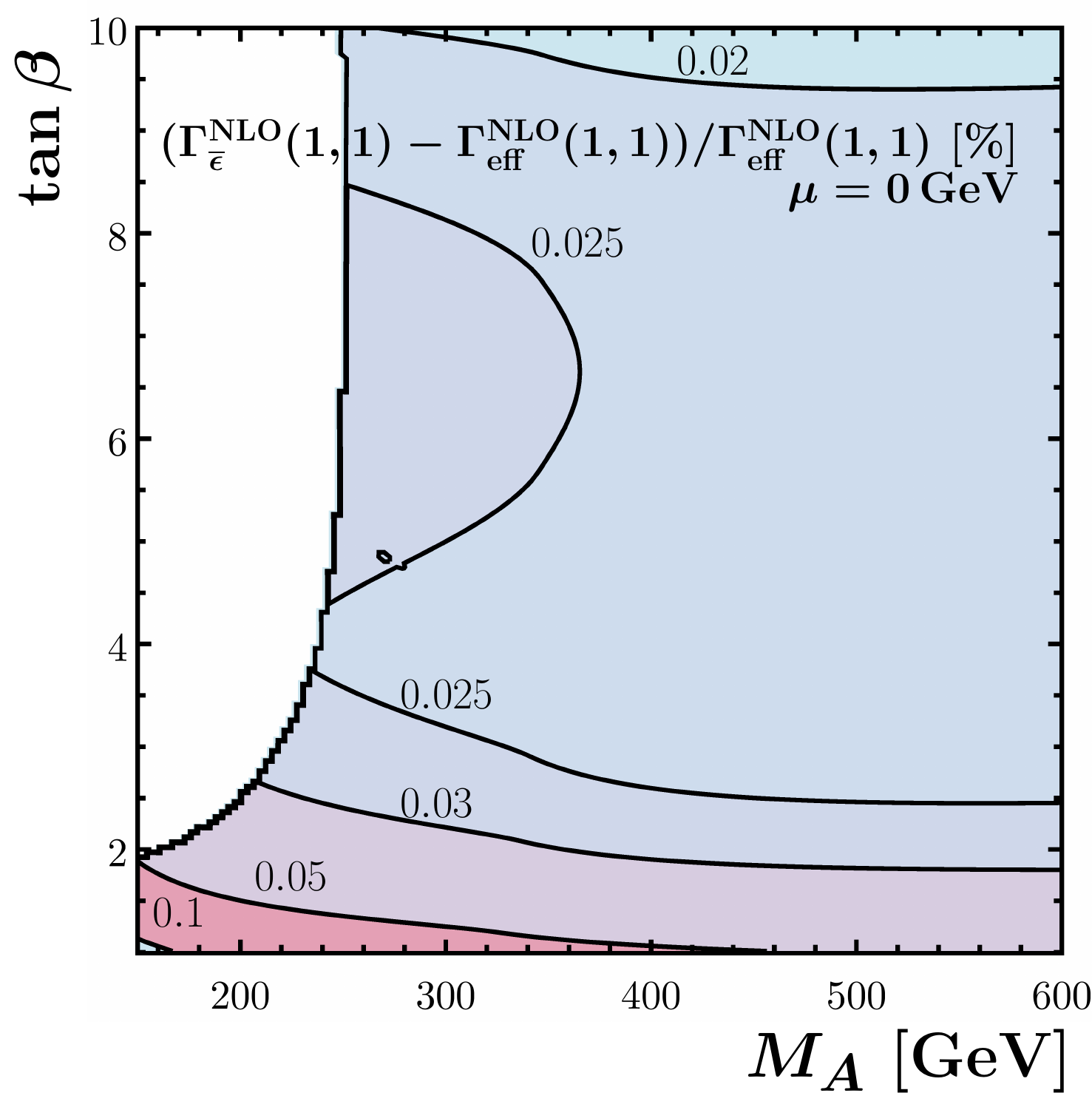} &
      \includegraphics[width=0.3\textwidth]{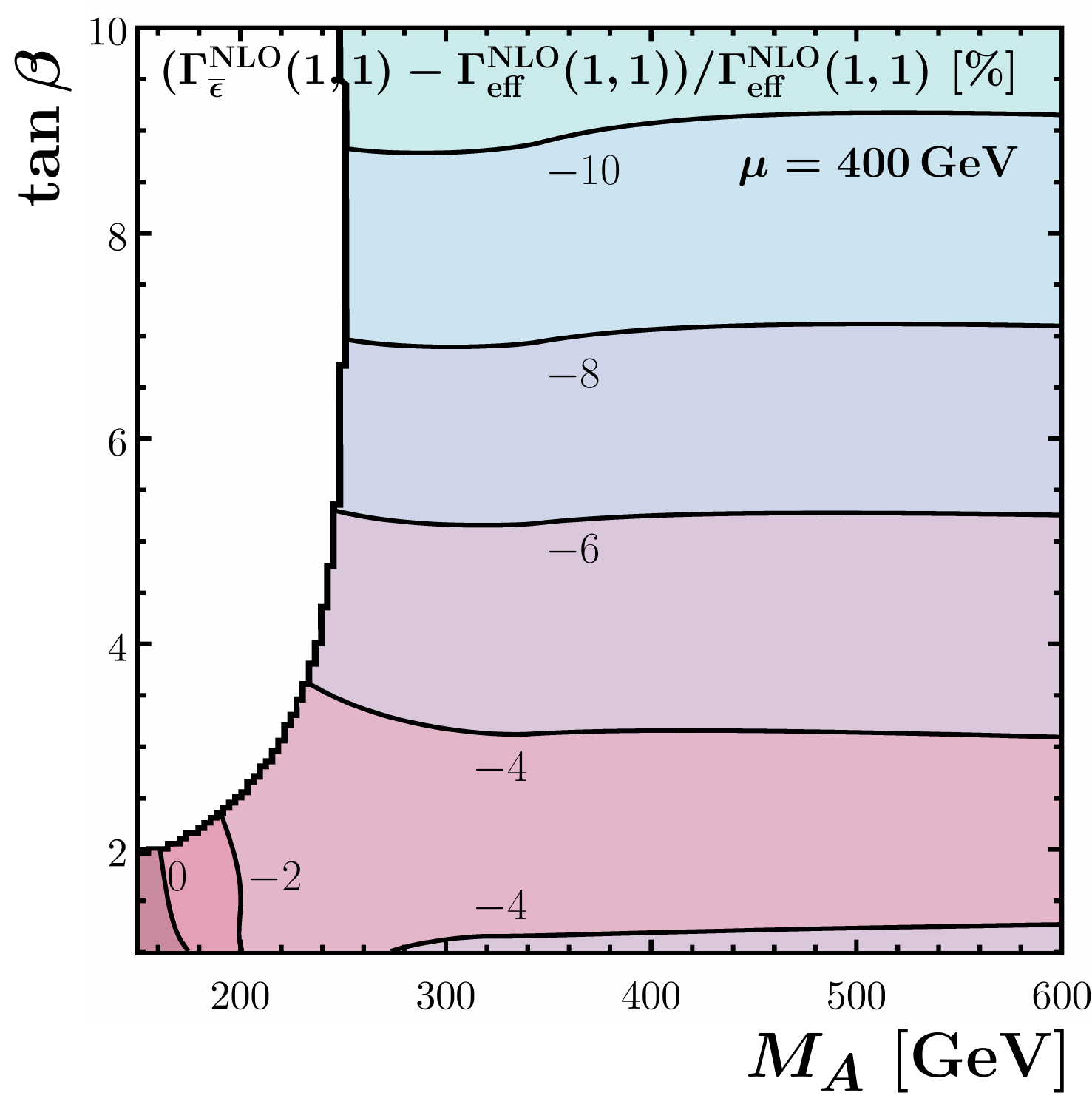} &
      \includegraphics[width=0.3\textwidth]{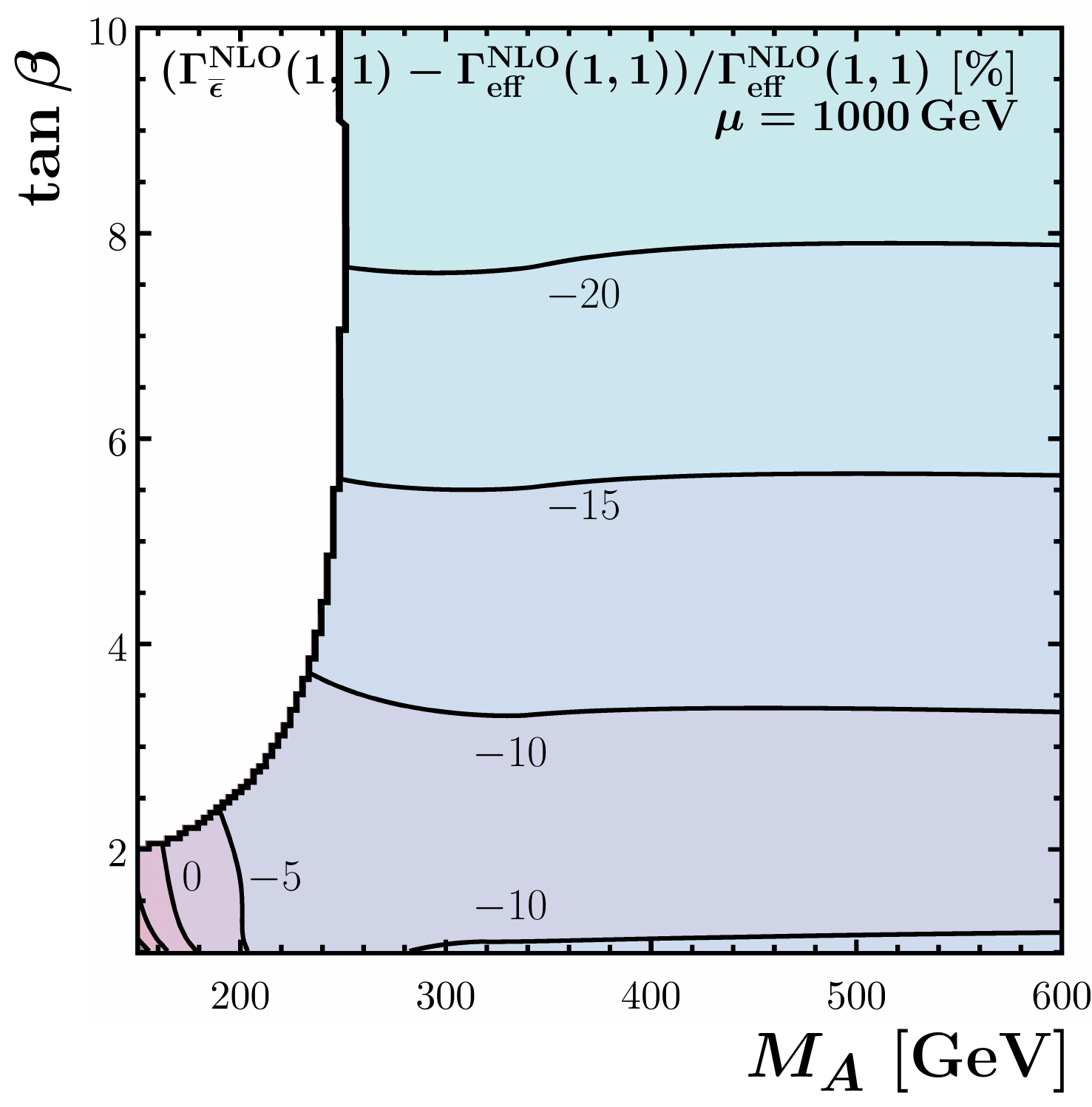}
      \\[-0.5cm]
      (d) & (e) & (f)
    \end{tabular}
    \parbox{\textwidth}{
      \caption[]{\label{fig:Gammaepsilon} (a,b,c) $\Gamma^{\text{\nlo{}}}_{\epsilon}(1,1)$ and (d,e,f) $\Gamma^{\text{\nlo{}}}_{\overline\epsilon}(1,1)$
      relative to $\Gamma^{\text{\nlo{}}}_{\text{eff}}(1,1)$
      as a function of $M_A$ in GeV and $\tan\beta$ for (a,d) $\mu=0$\,GeV, (b,e) $\mu=400$\,GeV and (c,f) $\mu=1000$\,GeV.}}
  \end{center}
\end{figure}

We finally focus on the performance of the \hmssm{} approximation in the description
of the partial decay width $H\to hh$.
In \fig{fig:Gammaepsilon} we compare the partial decay widths $\Gamma^{\text{\nlo{}}}_{\epsilon}(1,1)$
and $\Gamma^{\text{\nlo{}}}_{\overline\epsilon}(1,1)$, both named option 2,
against the full result $\Gamma^{\text{\nlo{}}}_{\text{eff}}(1,1)$, named option 3 in \sct{sec:combresult}.
In a first comparison we keep both top-quark and stop corrections in our calculation. For $\mu=0$\,GeV 
it is apparent, see \fig{fig:Gammaepsilon}~(a), that $\Gamma^{\text{\nlo{}}}_{\epsilon}(1,1)$ deviates from the exact
result by more than $20$\% for small values of $\tan\beta$. Using instead the improved Higgs self-coupling
$\lambda^{\overline\epsilon}_{Hhh}$ in $\Gamma^{\text{\nlo{}}}_{\bar{\epsilon}}(1,1)$ results in almost perfect agreement,
see \fig{fig:Gammaepsilon}~(d) with tiny deviations of only maximally $0.1$\%. \s

Choosing a non-vanishing value of $\mu$, see \fig{fig:Gammaepsilon}~(b) and (c)
for $\mu=400$\,GeV and $\mu=1000$\,GeV, respectively, leads to deviations
between the \hmssm{} prediction $\Gamma^{\text{\nlo{}}}_{\epsilon}(1,1)$ and
the exact result $\Gamma^{\text{\nlo{}}}_{\text{eff}}$. Such
deviations increase with the value of $\mu$ 
and follow a similar pattern as observed for $\lambda^{\epsilon}_{Hhh}$
when compared against $\lambda^{\text{eff}}_{Hhh}$, see \fig{fig:lambda}.
Keep in mind, that the squared value of $\lambda$ enters the \lo{} partial decay width.
Again the usage of $\lambda^{\overline\epsilon}_{Hhh}$, depicted in \fig{fig:Gammaepsilon}~(e) and (f),
makes the overall spread of the discrepancies between the \hmssm{} prediction
and the exact result smaller, but in particular for larger values of $\tan\beta$
they can still exceed $10$\,\%. \s

\begin{figure}
  \begin{center}
    \begin{tabular}{ccc}
      \includegraphics[width=0.3\textwidth]{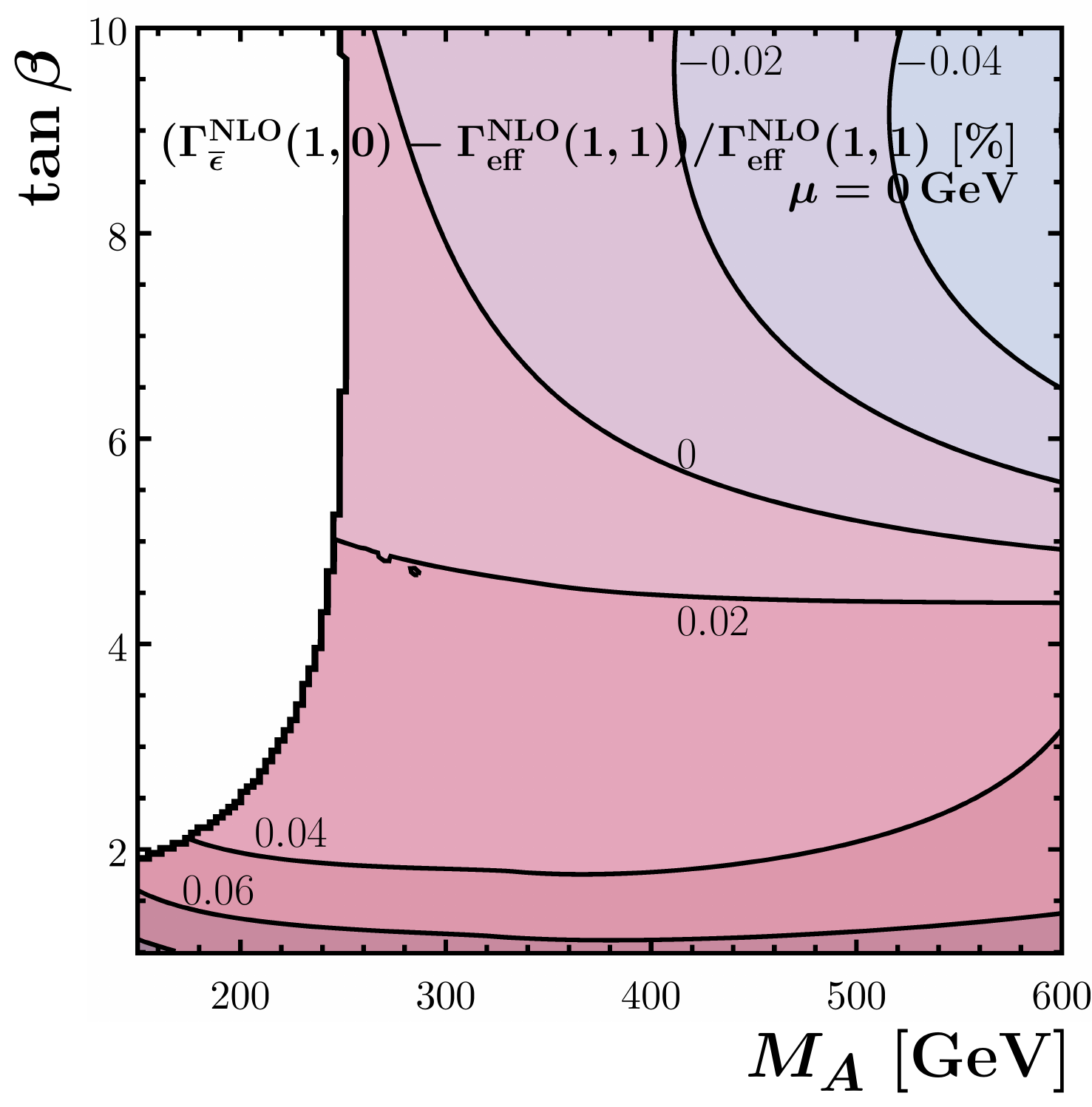} &
      \includegraphics[width=0.3\textwidth]{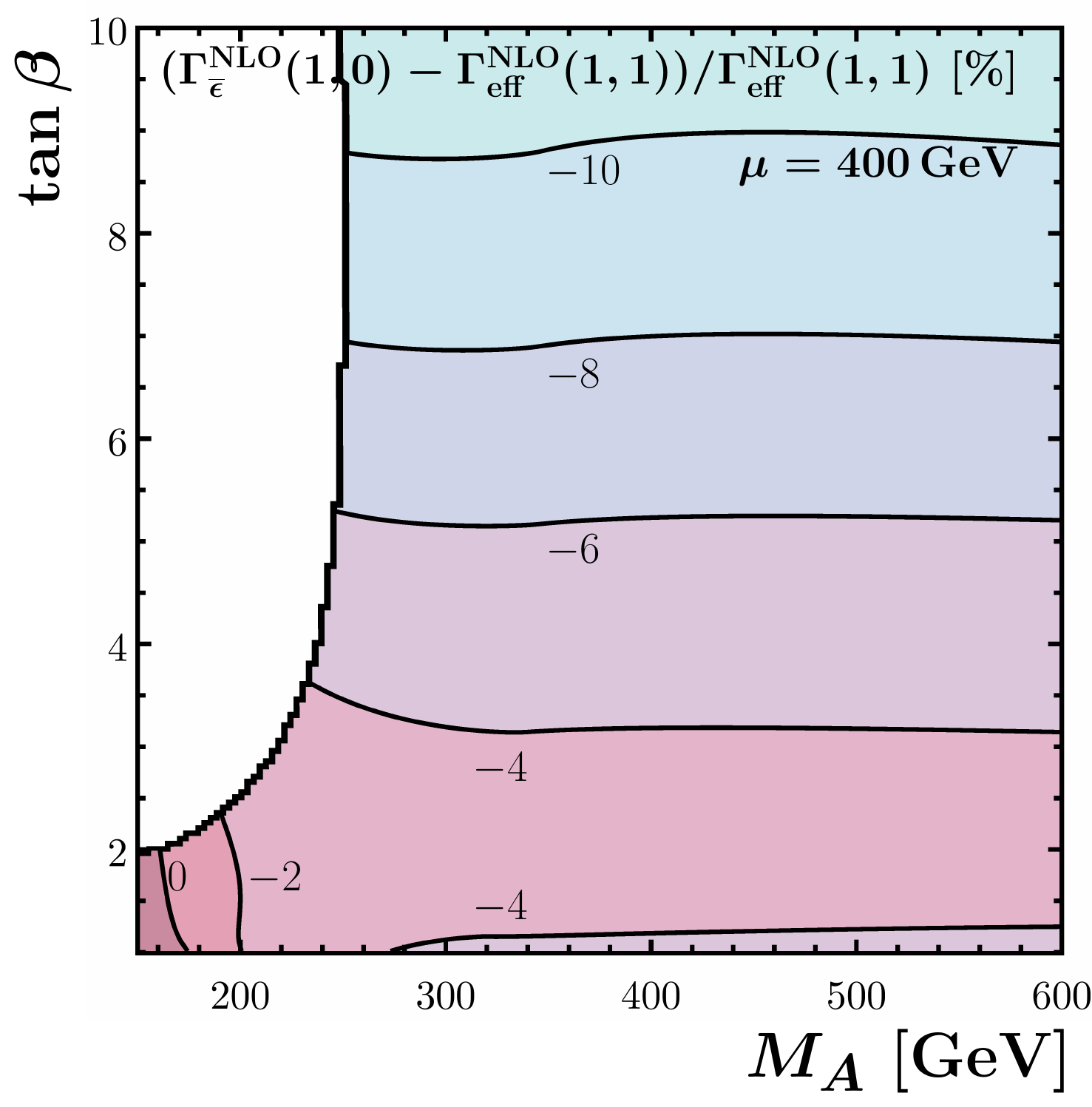} &
      \includegraphics[width=0.3\textwidth]{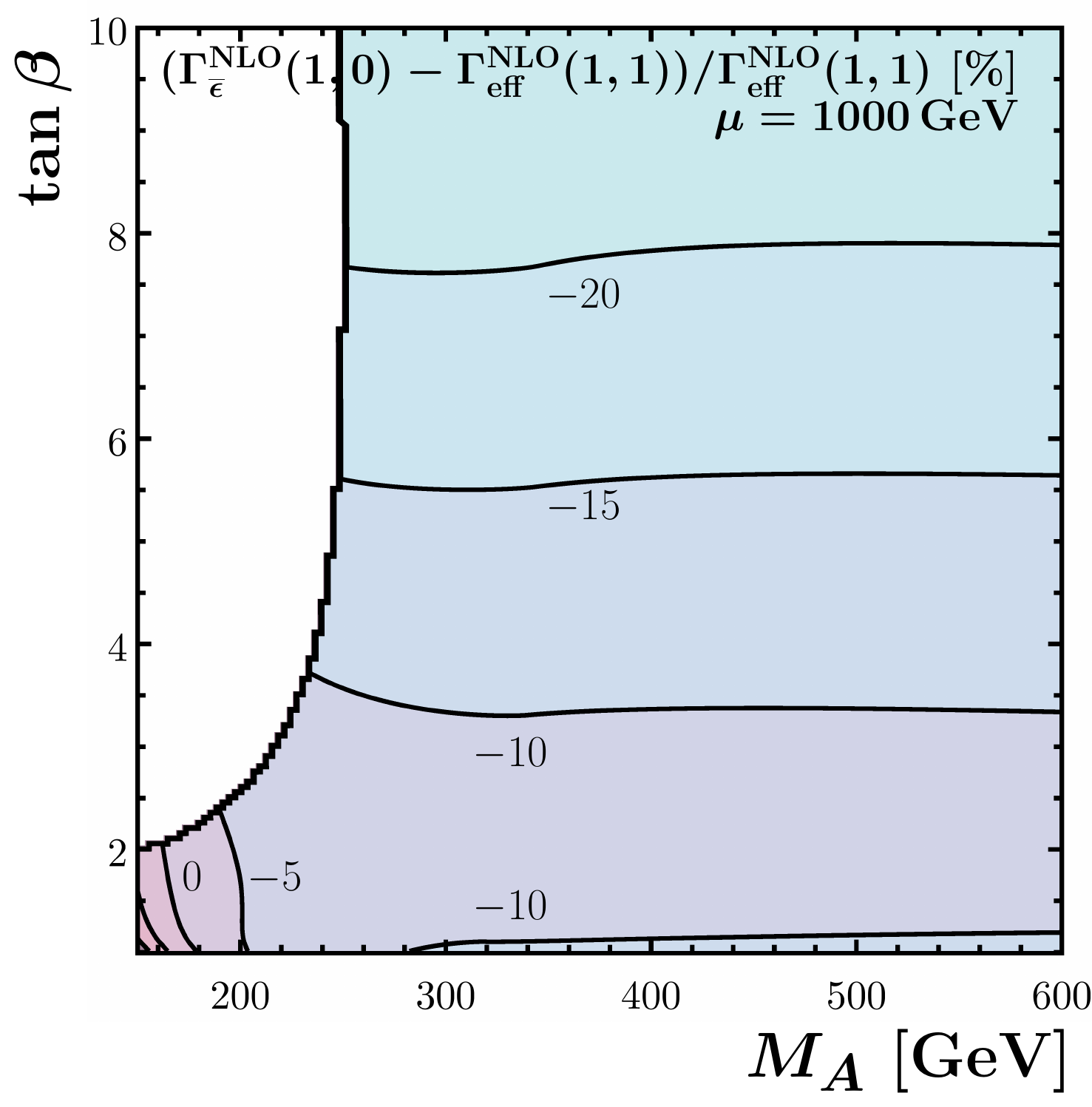}      
      \\[-0.5cm]
      (a) & (b) & (c)
    \end{tabular}
    \parbox{\textwidth}{
      \caption[]{\label{fig:Gammaepsilon2} (a,b,c) $\Gamma^{\text{\nlo{}}}_{\overline\epsilon}(1,0)$
      relative to $\Gamma^{\text{\nlo{}}}_{\text{eff}}(1,1)$
      as a function of $M_A$ in GeV and $\tan\beta$ for (a) $\mu=0$\,GeV, (b) $\mu=400$\,GeV and (c) $\mu=1000$\,GeV.}}
  \end{center}
\end{figure}

Lastly, we suggest to calculate $\Gamma^{\text{\nlo{}}}_{\overline\epsilon}(1,0)$ in the \hmssm{} approach,
since this calculation of the decay width incorporates the most dominant one-loop corrections from top quarks,
but on the other hand is not sensitive to the actual supersymmetric spectrum and is thus in the spirit
of the \hmssm{} approach.
This decay width is compared against the exact result including top-quark and stop corrections in \fig{fig:Gammaepsilon2}.
From the fact that for small $\mu$ values there is only a small difference to the results including one-loop
stop corrections $\Gamma^{\text{\nlo{}}}_{\overline\epsilon}(1,1)$, compare e.g. \fig{fig:Gammaepsilon}~(d)
against \fig{fig:Gammaepsilon2}~(a), we again conclude that momentum-dependent stop corrections are subdominant,
at least for our benchmark scenario with stop masses above $1$\,TeV and a heavy Higgs mass $m_H$
well below the TeV scale. The comparison of \fig{fig:Gammaepsilon}~(e) and (f)
against \fig{fig:Gammaepsilon2}~(b) and (c), respectively, shows that also
for larger~$\mu$ the momentum-dependent corrections
from the stop sector can well be neglected.

\section{Conclusions \label{sec:concl}}

We revisited the calculation of the decay of the heavy Higgs boson~$H$ into
two \sm{}-like Higgs bosons~$h$ in the \hmssm{} approach.
For this purpose we considered the full effective potential,
in which the top quark and stops are integrated out,
allowing to match the \mssm{} on a \thdm{} as its low-energy limit.
By carefully integrating out the top quark and stops separately,
we identified missing contributions in the Higgs self-couplings
of the \hmssm{} approach and suggested the definition of
improved Higgs self-couplings for the \hmssm{} approximation.
In particular in the limit of a small Higgsino mass parameter $\mu\ll M_S$,
which is an intrinsic assumption
of the \hmssm{} approach, these improved
Higgs self-couplings yield an excellent approximation for
the actual Higgs self-couplings calculated in our effective potential approach.
In this context the \hmssm{} approach can and has to
be understood as an approximation to a full effective low-energy \thdm{}
matched to the \mssm{}. Since the effective
potential is evaluated for zero external momenta,
we included kinetic corrections at the level of the effective
Lagrangian of the \thdm{} involving dimension-$4$ operators.
These corrections allow us to perform a one-loop Feynman-diagrammatic calculation
of $H\to hh$ in the \thdm{}-setup, where again we can consider
the top quark and stops separately.
By this procedure we discussed the relevance of momentum-dependent and kinetic
corrections to the decay $H\to hh$ at the one-loop level, that
were missing in previous comparisons.
As for the Higgs self-couplings themselves
using our improved Higgs self-couplings for the decay $H\to hh$ leads
to a much better agreement between the calculation performed in
the \thdm{}-setup and the \hmssm{} approximation, in particular for small $\mu \ll M_S$.
We advocate to supplement the calculation of $H\to hh$ with
momentum-dependent and kinetic corrections from the top quark to obtain a reliable
prediction at one-loop level without introducing a dependence
on the actual supersymmetric particle spectrum.
Our study misses an inclusion of RGE effects, such that also heavier supersymmetric
spectra can be investigated. We leave such comparisons to future work.

\section*{Acknowledgements}
This work was inspired by comparisons performed in the context
of the LHC Higgs XS working group. We thank Pietro Slavich for discussions.
S.L. and M.M.  are supported by the DFG project
``Precision Calculations in the Higgs Sector - Paving the Way to the New Physics Landscape'' (ID: MU 3138/1-1).

\appendix
\section{Analytical results}
\label{app:analyticresults}

\begin{figure}[h]
  \begin{center}
    \begin{tabular}{cc}
      \parbox{0.2\textwidth}{\includegraphics[width=0.2\textwidth]{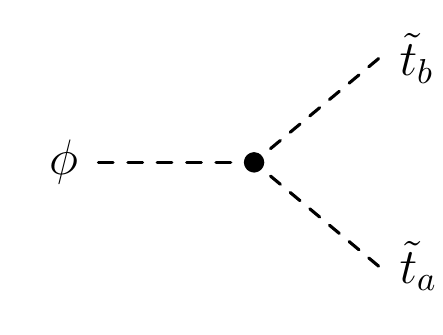}}$=-i2\frac{G_{ab}^\phi}{v}\delta_{ij}$ &
      \parbox{0.2\textwidth}{\includegraphics[width=0.2\textwidth]{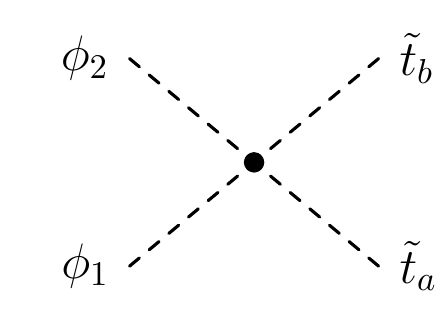}}$=-i2\frac{F_{ab}^{\phi_1\phi_2}}{v^2}\delta_{ij}$
    \end{tabular}
    \parbox{\textwidth}{
      \caption[]{\label{fig:feynmanrules} Feynman rules for interactions of squarks and Higgs bosons.
      The indices $i$ and $j$ refer to color, the indices $a$ and $b$ to mass eigenstates.}}
  \end{center}
\end{figure}

We subsequently present the relevant analytical results, being
the self-energies of the Higgs bosons and the virtual corrections
to the $Hhh$ vertex. We make use of Passarino-Veltman integrals \cite{tHooft:1978jhc, Passarino:1978jh}
to simplify our notation.
For their presentation we need the Feynman rules for squark to Higgs-boson couplings, which we depict
in \fig{fig:feynmanrules}. Note again that we work in the gaugeless limit, such
that $D$-term contributions to these couplings are neglected consistently.
The couplings for the three-point interaction in
\fig{fig:feynmanrules} are given by ($\phi \in \{h,H\}$)
\begin{align}
 G_{11/22}^\phi &=m_t^2g_t^\phi\pm g_{LR}^\phi s_{2\vartheta}\,,&\qquad
 &G_{12}^\phi = g_{LR}^\phi c_{2\vartheta} \,,&\\
 g_{LR}^h&=\frac{m_t}{2}\left(A_t g_t^h+\mu g_t^H\right)\,,&\qquad
 &g_{LR}^H=\frac{m_t}{2}\left(A_t g_t^H-\mu g_t^h\right) \,,&\\
 g_t^h &=\frac{c_\alpha}{s_\beta}\,,&\qquad 
 &g_t^H=\frac{s_\alpha}{s_\beta}\,,&
\end{align}
and for the four-point interaction by ($\phi_{1,2} \in \{h,H\}$)
\begin{align}
 F_{ab}^{\phi_1\phi_2}=m_t^2g_t^{\phi_1}g_t^{\phi_2}\delta_{ab}\,.
\end{align}
Therein $\vartheta$ denotes the squark-mixing angle that
diagonalizes the squark mass matrix in \eqn{eq:squarkmassmatrix}.
Using $C_t$ from \eqn{eq:ctabb} the mixing angle can also be reexpressed
as $c_{2\vartheta}^2=1-4m_t^2C_t^2$.

\subsection{Vertex corrections}
For the Feynman diagram involving the top-quark correction to $H\to hh$, depicted in \fig{fig:feynman}~(b),
we obtain
\begin{align}\nonumber
 \mathcal{A}_1=-i\frac{m_t^4}{v^3}g_t^H(g_t^h)^2\frac{8N_c}{(4\pi)^2}&\left[B_0(q_H^2;m_t^2,m_t^2)+B_0(q_1^2;m_t^2,m_t^2)+B_0(q_2^2;m_t^2,m_t^2)\right.\\
 &\left.+\left(4m_t^2-\frac{q_H^2+q_1^2+q_2^2}{2}\right)C_0(q_H^2,q_1^2,q_2^2;m_t^2,m_t^2,m_t^2)\right]\,,
\end{align}
where $N_c=3$ and we fix external momenta to the on-shell masses, i.e. $q_H^2=M_H^2, q_1^2=q_2^2=M_h^2$.
For the squark contributions we split the contribution from the Feynman diagram in \fig{fig:feynman}~(c)
from the sum of the contributions from the diagrams in \fig{fig:feynman}~(d)--(f). For the former we obtain
\begin{align}
 \mathcal{A}_2=i\sum_{a,b,c}\frac{16N_c }{(4\pi)^2v^3}G_{ab}^HG_{ac}^hG_{bc}^hC_0(q_H^2,q_1^2,q_2^2;\msta^2,\mstb^2,\mstc^2)\,,
\end{align}
and the sum of the diagrams in \fig{fig:feynman}~(d)--(f) yields
\begin{align}
 \mathcal{A}_3=i\sum_{a,b}\frac{4N_c}{(4\pi)^2v^3}\left[ G_{ab}^HF_{ab}^{hh}B_0(q_H^2;\msta^2,\mstb^2)+2G_{ab}^hF_{ab}^{hH}B_0(q_1^2;\msta^2,\mstb^2)\right]\,,
\end{align}
where again external momenta are set on-shell.
It is clear that $\mathcal{A}^{\text{virt}}(1,1)$ is obtained by summing $\sum_{i=1,2,3}\mathcal{A}_i$,
while $\mathcal{A}^{\text{virt}}(1,0)$ only consists of $\mathcal{A}_1$.

\subsection{Self-energy corrections}
Again we start with the top-quark contribution, which is depicted in \fig{fig:feynman}~(l) and is given by
\begin{align}
 \Sigma^t_{\phi_i\phi_j}(p^2)&=\frac{m_t^2}{v^2}g_t^{\phi_i}g_t^{\phi_j}\frac{2N_c}{(4\pi)^2}\left[2A_0(m_t^2)+(4m_t^2-p^2)B_0(p^2;m_t^2,m_t^2)\right]\,,\\
 \Sigma'^t_{\phi_i\phi_i}(p^2)&=\frac{m_t^2}{v^2}(g_t^{\phi_i})^2\frac{2N_c}{(4\pi)^2}\left[-B_0(p^2;m_t^2,m_t^2)+(4m_t^2-p^2)B'_0(p^2;m_t^2,m_t^2)\right]\,,
\end{align}
where we attach two Higgs bosons $\phi_i,\phi_j\in\lbrace h,H\rbrace$ externally.
Finally for the sum of all stop contributions, shown in
\fig{fig:feynman}~(m, n), we obtain
\begin{align}
 \Sigma^{\tilde t}_{\phi_i\phi_j}(p^2)&=-\frac{2N_c}{(4\pi)^2v^2}\left[\sum_a F_{aa}^{\phi_i\phi_j}A_0(\msta^2)+2\sum_{a,b}G_{ab}^{\phi_i}G_{ab}^{\phi_j}B_0(p^2;\msta^2,\mstb^2)\right]\,,\\
 \Sigma'^{\tilde t}_{\phi_i\phi_j}(p^2)&=-\frac{4N_c}{(4\pi)^2v^2}\sum_{a,b}G_{ab}^{\phi_i}G_{ab}^{\phi_j}B'_0(p^2;\msta^2,\mstb^2)\,.
\end{align}
Again, $\mathcal{A}^{\text{ext}}(t,\tilde t)$ and $\mathcal{A}^{\text{ext,eff}}(t,\tilde t)$ defined in \eqn{eq:Aext} and \eqn{eq:Aexteff}, respectively,
contain for $(t,\tilde t)=(1,0)$ only contributions from $\Sigma^t$ and $\Sigma'^t$ and for $(t,\tilde t)=(1,1)$ in addition from $\Sigma^{\tilde t}$ and $\Sigma'^{\tilde t}$.

{\footnotesize
\bibliographystyle{utphys}
\bibliography{Hhh_hMSSM}
}

\end{document}